\documentclass[
reprint,
 aps,
 superscriptaddress,
bibnotes,
]{revtex4-2}

\usepackage{physics}
\usepackage{dsfont}
\usepackage{color,newfloat}
\usepackage{graphicx}
\usepackage{dcolumn}
\usepackage{bm}
\usepackage{amsmath,amssymb,amsthm}
\usepackage{mathtools}
\usepackage{multirow}
\usepackage[dvipsnames]{xcolor}
\usepackage{hyperref}
\hypersetup{
    colorlinks=true,
    linkcolor=Maroon,
    citecolor=Maroon,
    urlcolor=Maroon
}
\usepackage[caption = false]{subfig}

\usepackage{tikz}
\usetikzlibrary{calc}

\usepackage{chngcntr} 

\usepackage{enumerate}

\usepackage[ruled,vlined,linesnumbered]{algorithm2e}
\SetKwComment{Comment}{$\triangleright$\ }{}

\begin{document}


\title{
Optimising graph codes for measurement-based loss tolerance
}

\author{Tom J. Bell}
\email{tj.bell@bristol.ac.uk}
\affiliation{Quantum Engineering Technology Labs, H. H. Wills Physics Laboratory and Department of Electrical and Electronic Engineering, University of Bristol, BS8 1FD, UK.}
\affiliation{Quantum Engineering Centre for Doctoral Training, University of Bristol, UK.}

\author{Love A. Pettersson}
\affiliation{Center for Hybrid Quantum Networks (Hy-Q), Niels Bohr Institute, University of Copenhagen, Blegdamsvej 17, DK-2100 Copenhagen, Denmark.}

\author{Stefano Paesani}
\email{stefano.paesani@nbi.ku.dk}
\affiliation{Center for Hybrid Quantum Networks (Hy-Q), Niels Bohr Institute, University of Copenhagen, Blegdamsvej 17, DK-2100 Copenhagen, Denmark.}


\begin{abstract}

Graph codes play an important role in photonic quantum technologies as they provide significant protection against qubit loss, a dominant noise mechanism. 
Here, 
we develop 
methods 
to analyse and optimise
measurement-based tolerance to 
qubit loss 
and computational errors
for arbitrary graph codes. 
Using these tools we identify 
optimised codes
with up to 12 qubits and
asymptotically-large modular constructions.
The developed methods
enable significant benefits
for various photonic quantum technologies, as  we illustrate 
with novel all-photonic quantum repeater states 
for quantum communication 
and high-threshold fusion-based schemes for fault-tolerant quantum computing. 

\end{abstract}

\date{\today}

\maketitle



\begin{figure}[]
  \centering
  \includegraphics[
  width=0.45 \textwidth]{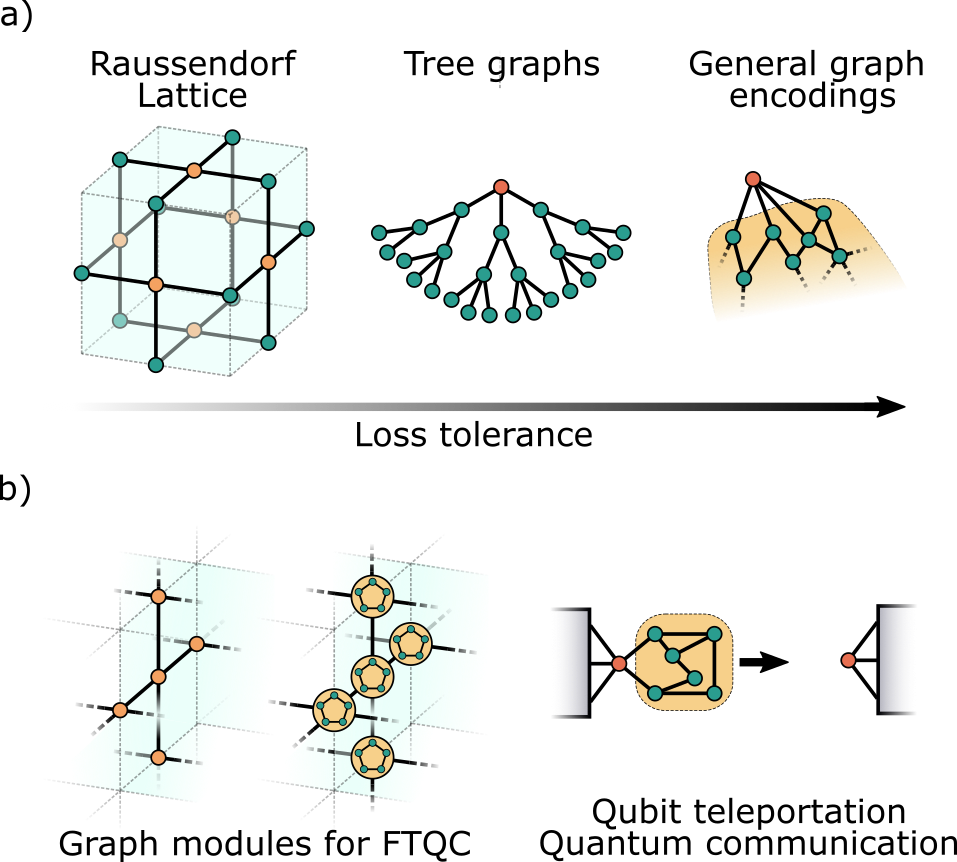}
  \caption{
  a) Graph codes can encode quantum information with inherent robustness to qubit loss. 
  Example of loss-tolerant graphs include the Raussendorf lattice~\cite{raussendorf2007} and tree graphs~\cite{Varnava2006}.
  In this work we develop tools to analyse the loss tolerance for arbitrary graph codes. 
  b) Graph codes with improved performance can enable designing better modules for modular photonic applications, including computation or communication schemes.}
  \label{fig:story}
\end{figure}

%
Quantum information is fragile and its control can be easily impaired by dissipation in the physical environment.
Quantum error correction (QEC) and fault tolerance aim at reducing the impact of noise, provided the physical error rate is below a certain threshold, enabling the control of quantum information in spite of physical imperfections~\cite{Shor96, gottesman1997, Knill1998}.
Optimising codes to the specific platform used and targeting the native noise mechanisms and operations to reduce the operational overheads is key to make QEC practical for near-term and future quantum hardware. 
In photonics, the dominant noise mechanism is photon loss, which irreversibly erases the state of the associated physical qubit. 
Although photon loss is an error that can be directly detected, unlike conventional gate errors, it nevertheless poses stringent hardware requirements for practical applications.
For example, current architectures for fault-tolerant photonic quantum computing need losses to be below a threshold of approximately $2\%$~\cite{fbqc}, very challenging for photonic set-ups.
A possible modular approach to improve these requirements is to encode each computational qubit in a loss-tolerant code, as pictured in Fig.~\ref{fig:story}.
The encoding and decoding are typically measurement-based, i.e. obtained by sequential destructive measurements on part of an entangled resource state to protect the remaining unmeasured components from errors -- an approach particularly suitable for photonics~\cite{Raussendorf03, hein2006, rudolph2017, fbqc}.
As we will show in this work, codes with moderate size, less than a few tens of qubits, can already provide significant measurement-based suppression of logical errors due to photon loss on encoded qubits.
Previous proposals have considered various types of loss-tolerant codes, e.g. tree graph codes~\cite{Varnava2006} (see Fig.~\href{fig:story}{\ref{fig:story}a}) and Beacon-Shor codes~\cite{Shor95}, whose code structures allow the loss tolerance to be readily analysed.
The use of these codes was proposed and investigated, for example, in the context of photonic measurement-based quantum communication~\cite{Azuma2015, Borregaard2020, Zhan2022repeaterperf, niu2022all}  and computation~\cite{fbqc, li2022concatenation}.
Identifying resource-efficient codes with high loss tolerance could bring significant practical benefits to these technologies.
Here, we address this goal by developing methods to analyse the loss and error tolerance in general graph codes and use them to design and implement optimisation techniques.
We fully characterise measurement-based fault-tolerant properties and optimise graph codes with up to 12 qubits, and investigate generalisations to larger graphs with modular structures.
We find optimised codes that can provide significant advantages in various photonic applications, including improved repeater graph states for quantum communication and fusion-based schemes for fault-tolerant photonic quantum computing with loss thresholds up to $10.5\%$ using standard linear optical fusions.


\section{Graph codes}
\label{sec:background}
\subsection{Encoding a logical qubit}
\label{sec:graph states}

The codewords of graph codes are graph states, a class of quantum states that can be conveniently described in terms of graphs and graph transformations~\cite{Hein2004PRA, hein2006}. %
The quantum state associated to an undirected and unweighted graph $G=(V, E)$ with vertices $V$ and edges $E$ is
\begin{equation}
\label{eq::graph_state_def}
    \ket{G} = \prod_{(i, j)\in E} CZ_{i,j} \ket{+}^{\otimes \mid V \mid},
\end{equation}
where $CZ_{i,j}$ represents a controlled-$Z$ operation between qubits $i$ and $j$, and is represented by edge between the associated vertices. 
Graph states are stabilizer states~\cite{gottesman1997} with stabilizer generators $K_{i} = X_{i}\prod_{k\in\mathcal{N}_{i}}Z_{k}$, where $i$ runs over the graph nodes and $\mathcal{N}_{i}$ is the neighbourhood of qubit $i$.
Throughout this work, we use $X$, $Y$, $Z$ to indicate Pauli operators. 
The generators generate the Abelian group of stabilizer operators $\mathcal{S} = \langle K_{j} \rangle_{j=1}^{n}$, meaning that each stabilizer $S\in\mathcal{S}$ is a product of generators $S = \prod_{i=1}^{n}K_{i}^{b_{i}}$, with $b_{i} \in \{0, 1\}$. 

\begin{figure*}
    \centering
    \includegraphics[
  width=1.0 \textwidth]{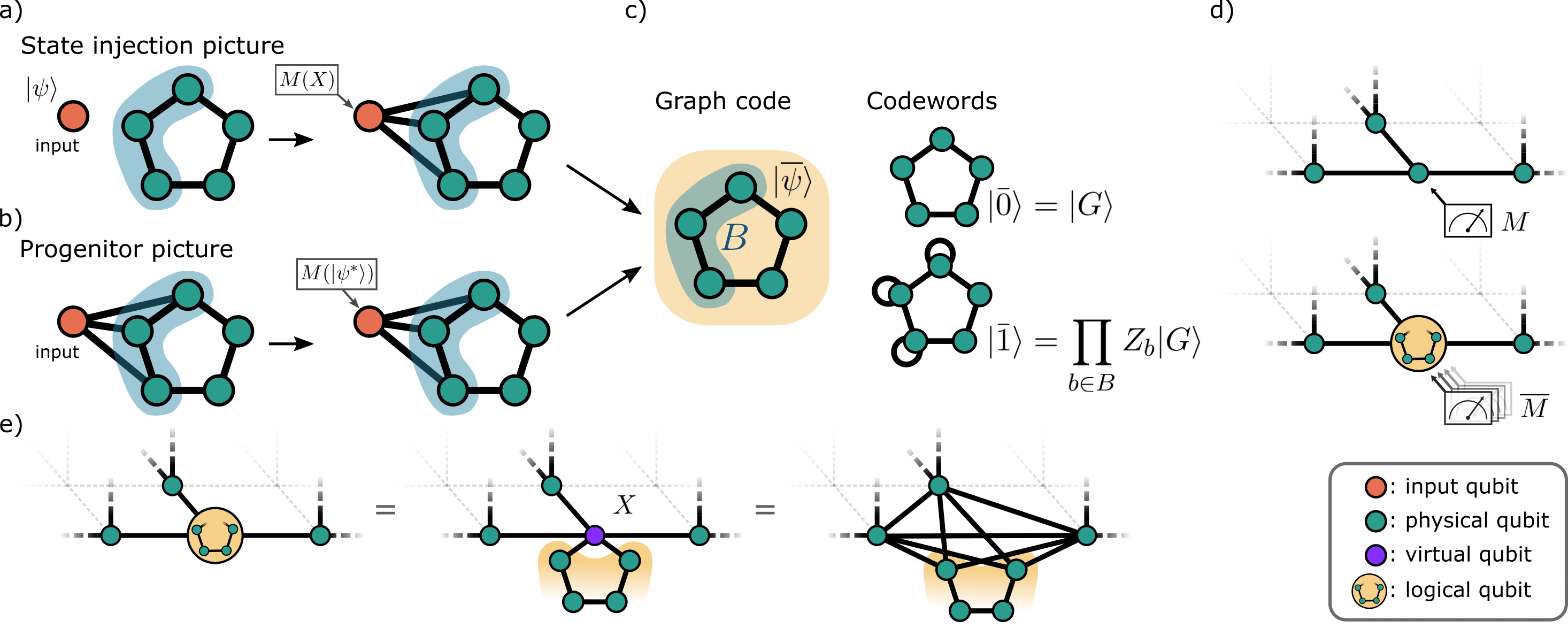}
    \caption{
    \textbf{Encoding in graph states.} 
    a) State injection picture, where the logical state is encoded by preparing it on an ancilla qubit, followed by entangling operations and an $X$ measurement on the ancilla. 
    b) Progenitor picture, where the same encoding is performed measuring the input qubit of the progenitor graph in the targeted state.
    c) The prepared graph code in the logical state $\ket{\overline{\psi}}$, and associated codewords. 
    Loop-edges represent $Z$ Pauli operations on the associated qubits.
    d) Modular schemes for measurement-based operations, where computational qubits are encoded in in graph codes and computational qubit measurements correspond to logical measurements on the graph codes.
    e) Physical modularisation of resource states, obtained substituting each computational qubits with a virtual qubit, the input of the associated graph code, to be measured in $X$.}
    \label{fig:graph_codes}
\end{figure*}

A possible method to encode an arbitrary qubit state $\ket{\psi}=\alpha\ket{0} + \beta\ket{1}$ in an $n$-qubit graph code is pictured in Fig.~\href{fig:graph_codes}{\ref{fig:graph_codes}a}. 
We initially prepare the $n$ physical code qubits in a graph state $\ket{G}$ and an additional input qubit in the target state $\ket{\psi}$.
The logical encoding is then performed by applying controlled-$Z$ operations between the input qubit and a subset $B$ of the code qubits, and then measuring the input qubit in the $X$-basis.
As these operations are all in the Clifford group, we can track the stabilizers and logical operations on the state according to the usual stabilizer transformation rules~\cite{gottesman1997}, obtaining the $n$-qubit code in the logical state $\ket{\overline{\psi}}=\alpha\ket{\overline{0}} + \beta\ket{\overline{1}}$ if the measurement outcome is $+1$, and $\overline{Z}\ket{\overline{\psi}}$ on outcome $-1$.
Here $\ket{\overline{0}}= \ket{G}$ and the logical operators are given by $\overline{X}= \prod_{b \in B}Z_{b}$ and $\overline{Z} = K_{b_0}$ for some choice of $b_0\in B$ (see  Fig.~\href{fig:graph_codes}{\ref{fig:graph_codes}c}). 
The updated code stabilizers are $\mathcal{S}=\langle K_{b_0}K_{b}, K_{b'}\rangle_{b\in B\backslash b_0, b' \notin B}$, with $K_{i}$ the stabilizer generators of $\ket{G}$~\cite{hein2006}.
As products of a logical operator with stabilizers also form valid logical operators of the code, we can write the set of all logical operators as $\overline{\mathcal{L}}=( \overline{X}, \overline{Z}, \overline{Y}= i\overline{X}\,\overline{Z}  ) \cdot \mathcal{S}$.
Note that an equivalent approach to  encoding can be obtained by initialising the input qubit in $\ket{+}$ before applying the controlled-$Z$ operations to the code qubits in $B$, and then measuring it in the qubit basis $\{\ket{\psi^{*}}, \ket{\psi^{*}_{\perp}}\}$ (which in general can be non-Clifford), as pictured in Fig.~\href{fig:graph_codes}{\ref{fig:graph_codes}b}.
In this way, the encoding can be described as starting from a fixed \textit{progenitor} graph $G'=(V+\{\text{input}\}, E+(\{\text{input}\}, B))$ with $n+1$ nodes, and then measuring the input node in the targeted basis. 
This can be simply observed by writing the initial graph state as $\ket{G'} = \left(\ket{0}_{\text{in}}\ket{G} + \ket{1}_{\text{in}}\prod_{b\in B}Z_{b}\ket{G}\right)/\sqrt{2} = \left(\ket{0}_{\text{in}}\ket{\overline{0}} + \ket{1}_{\text{in}}\ket{\overline{1}}\right)/\sqrt{2}$ with codewords and logical operators defined as above, for which projective measurement of the input qubit into $\ket{\psi^{*}}$ prepares the logical code state in the conjugate state $\ket{\overline{\psi}}$, as desired.
The two pictures are equivalent but, depending on the protocol under study, analysing the encoding in the progenitor picture may be more convenient as it permits to describe the encoding and decoding of a logical qubit entirely through the stabilizers of the progenitor graph.
In fact, the stabilizer generators of the progenitor graph that act as $X_{\text{in}}$ and $Z_{\text{in}}$ on the input qubit are transformed in the logical operators $\overline{X}$ and $\overline{Z}$ of the code, respectively, upon measurement of the input.
That is, $\overline{X} \leftrightarrow K'_{\text{in}}$ and  $\overline{Z} \leftrightarrow K'_{b_0}$, where $K'_i$ are the stabilizer generators of the progenitor graph $G'$ and $b_0\in B$.

\subsection{Measurement-based decoding}

Measurement-based approaches process logical quantum information by only performing destructive single-qubit measurements on the code qubits and classical feed-forward.
In the context of QEC, the constraints imposed by operating a single destructive measurement per qubit add significant limitations with respect to repeatedly performing parity checks in circuit-based approaches. 
Namely, given a single-qubit measurement pattern $M$ on the physical qubits, the only stabilizers that will be accessible are those that commute qubit-wise with $M$, i.e. the stabilizers in
\begin{equation}
    \mathcal{S}_M = \{S \in \mathcal{S}\  |\ [S_i, M_i]=0 \text{ for each qubit } i\},
    \label{eq:MBQEC_constraint}
\end{equation}
with $\mathcal{S}$ the initial code stabilizers.
Note that $\mathcal{S}_M \subseteq \mathcal{S}$ forms a stabilizer subgroup of $\mathcal{S}$ (Appendix~\ref{sec:reduced_stab}).
Therefore, the effect of these constraints is effectively to induce a reduced code $\mathcal{S}_M$ compatible qubit-wise with the measurement $M$. 
This can also be regarded, more abstractly, as a gauge-fixing procedure~\cite{Paetznick2013, Brown2020}.
Measurement-based decoding of gate errors, i.e. the inference of qubit errors from the measured syndromes, can then be performed equivalently as one would do in standard QEC by considering the reduced code $\mathcal{S}_M$ induced by $M$. 

\subsection{Effects of qubit loss}

The effect of qubit loss detected during measurements can be described similarly to the enforcement of a measurement pattern $M$ described above. 
If a qubit is lost, all stabilizers and logical operators that act non-trivially on that qubit are no longer measurable.
This enforces a qubit-wise constraint similar but stronger to Eq.~\ref{eq:MBQEC_constraint} as compatibility now requires an identity on a lost qubit rather than just a commuting operator. 
To maintain a concise notation when describing the effects of loss, we will write $M_i = \nexists$ to indicate that qubit $i$ was lost, and use the convention that $[A, \nexists]=0$ iff $A=\mathds{1}$. 
With this notation, we can write the set of stabilizers $\mathcal{S}_M$ compatible with $M$, which now includes also lost qubits, again exactly as Eq.~\ref{eq:MBQEC_constraint}. 
Also in this case $\mathcal{S}_M$ is a stabilizer subgroup of the initial stabilizer group $\mathcal{S}$; the presence of losses has the effect of reducing it further (see Appendix~\ref{sec:reduced_stab} for more details).
However, qubit losses also pose constraints on logical operators of the code as they cannot have support on a lost qubit.
These constraints can be included in a very similar way as for the stabilizers by writing the induced set of logical operators as
\begin{equation}
    \overline{\mathcal{L}}_M = \{\overline{L} \in \overline{\mathcal{L}}\  |\ [\overline{L}_i, M_i]=0 \text{ for each qubit } i\},
        \label{eq:MBQEC_constraint_logop}
\end{equation}
again using the convention $M_i = \nexists$ if qubit $i$ is lost. 
In the progenitor graph picture, the conditions in Eq.~\ref{eq:MBQEC_constraint_logop} can be conveniently included by directly applying Eq.~\ref{eq:MBQEC_constraint} to the stabilizers of the progenitor graph.
The main idea behind loss tolerant measurement-based QEC is that, if losses are not excessive, the set $\overline{\mathcal{L}}_M$ remains nontrivial and $\mathcal{S}_M$ contains enough stabilizers to protect the encoded logical state from errors.
In general, the code performance depends on the chosen single-qubit measurement pattern $M$, as well as on the initial graph code. 
If losses are \textit{heralded}, i.e. which qubits are lost is known before their measurement, the measurement pattern $M$ can be conveniently optimised beforehand to achieve the best available $\mathcal{S}_M$ and $\overline{\mathcal{L}}_M$~\cite{morley2019}.
However, losses are often \textit{unheralded}: loss of a qubit is detected only upon its measurement and not before.
This loss model is relevant to most quantum platforms (e.g. photonics), and we will focus on it in this work.
Finding initial graph codes and measurement strategies that provide good tolerance to unheralded loss is in general a complex task and will be investigated in the next sections.

\section{Loss-tolerant logical measurements with graph codes}
\label{sec::measurements}

In measurement-based approaches, computational operations are implemented via single-qubit measurements. 
We start describing general methods to perform these measurements loss-tolerantly when encoding each computational qubit into an arbitrary graph code, as depicted in Fig.~\href{fig:graph_codes}{\ref{fig:graph_codes}d-g}.

\subsection{Loss-tolerant logical Pauli measurements}

\begin{figure}
    \centering
    \includegraphics[
  width=0.45 \textwidth]{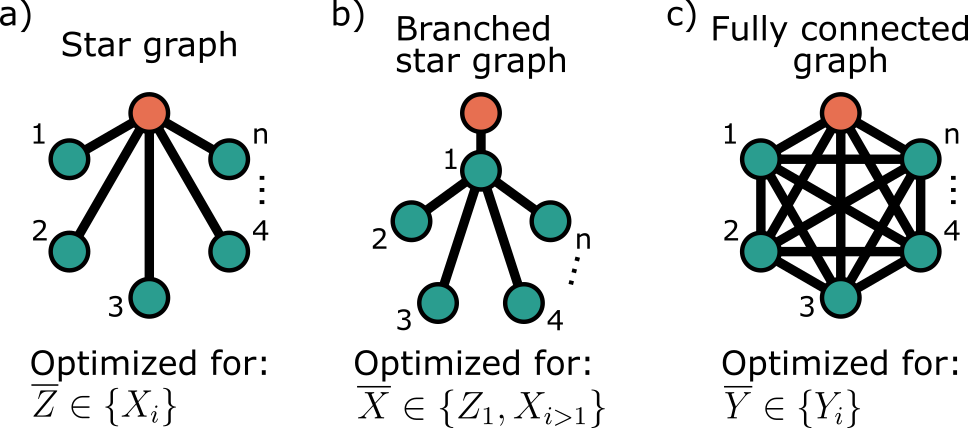}
    \caption{
    \textbf{Optimal graphs for individual logical Pauli measurements.} 
    a) Star graph providing optimal loss-tolerance for measuring $\overline{Z}$, b) branched star graph for $\overline{X}$, and c) star graph for $\overline{Y}$.
    Considering $n$ code qubits, for any of these graph structures the success probability to measure the associated logical operator is $1-\ell^n$, although for the remaining two logical Pauli operators it decreases exponentially as $(1-\ell)^n$, with $\ell$ the physical loss rate.}
    \label{fig:Paulimeas}
\end{figure}

As stabilizers are based on the Pauli group, logical measurements in the Pauli bases are the simplest to analyse within the framework we described: it corresponds to measuring a logical operator $\overline{L} \in \{\overline{X}, \overline{Y}, \overline{Z}\}$.
In the progenitor picture, this can be seen as a non-destructive Pauli measurement on the input qubit without having to directly measure it, often referred to as an \textit{indirect} measurement~\cite{Varnava2006}. 
Let us consider, for example, an indirect measurement of the $\overline{X}$ logical operator. 
In presence of qubit loss, a measurement pattern $M$ on the physical qubits successfully measures it if the set  $\overline{\mathcal{L}}{[X]}_M$, obtained applying the condition in Eq.~\ref{eq:MBQEC_constraint_logop} to the set $\overline{\mathcal{L}}{[X]}$ of all possible logical $\overline{X}$ operators of the code, is non-empty.
Equivalently, it means there exists a logical operator $\overline{X}\in \overline{\mathcal{L}}{[X]}_M$ that can be obtained from the single-qubit measurements performed in $M$ and with no support on lost qubits (i.e. $\overline{X}_i=\mathds{1}$ if $M_i = \nexists$).
Identical conditions apply for $\overline{Y}$ and $\overline{Z}$ indirect measurements. 
A simple example is the indirect logical $\overline{X}$ measurement on star-graph codes, i.e. graph codes with a star-graph progenitor (see Fig.~\href{fig:Paulimeas}{\ref{fig:Paulimeas}a}).
%
%
From the definitions in Section~\ref{sec:background}, it is evident that a single-qubit operator $X_i$ on any code qubit $i$ provides a valid logical $\overline{Z}$ operator, i.e. $\overline{\mathcal{L}}{[Z]} = \{  X_i\}_{i=1}^n$.
Therefore, choosing the measurement pattern $M=\prod_{i} X_i$, if at least one code qubit is not lost then $\overline{\mathcal{L}}{[Z]}_M$ is non-empty and the logical operator is successfully measured.  
The probability of failing to measure $\overline{Z}$, which we call logical loss, is thus $\overline{\ell}[Z] = \ell^n$ where $n$ is the number of physical code qubits and $\ell$ the qubit loss probability.
Such strong robustness for $\overline{Z}$ measurements for the star graphs comes at the expense of weak performance for $\overline{X}$ and $\overline{Y}$ measurements.
In fact, both $\overline{X}$ and $\overline{Y}$ are weight-$n$ operators, meaning that all physical qubits need to exist to obtain a successful logical measurement, so $\overline{\ell}[X] = \overline{\ell}[Y] = 1 - (1-\ell)^n$. 

\begin{algorithm}
    \caption{Pauli measurement decoder}\label{alg:Paulidecoder}
\KwIn{Set $\overline{\mathcal{L}}[A]$ of logical $A\in\{X, Y, Z\}$ code operators.}
\KwOut{Outcome of $\overline{A}$ if successful, \texttt{False} otherwise.}

Initialise the set of unmeasured qubit $\Theta=\{1,\ldots,n\}$ using all $n$ code qubits, and $M=\mathds{1}^{\otimes n}$\;

\While{$\Theta$ and $\overline{\mathcal{L}}[A]$ are non-empty}{
    $i, P_i \leftarrow \texttt{NextMeas}(\Theta, \overline{\mathcal{L}}[A])$. \Comment*[r]{Decide next qubit $i$ and Pauli operator $P_i$ to measure.}
    
    Measure qubit $i$, set $M_i= P_i$ if successful, $M_i= \nexists$ if qubit is lost\;  

    $\overline{\mathcal{L}}[A] \leftarrow \texttt{UpdateDecoder}(M, \overline{\mathcal{L}}[A])$\Comment*[r]{Update $\overline{\mathcal{L}}[A]$ according to Eq.\ref{eq:MBQEC_constraint_logop}.}
    
    Remove qubit $i$ from $\Theta$.
}
If $\overline{\mathcal{L}}[A]$ is non-empty return value of $\overline{A}\in\overline{\mathcal{L}}[A]$, else return \texttt{False}.
\end{algorithm}

\begin{figure*}
    \centering
    \includegraphics[  width=1 \textwidth]{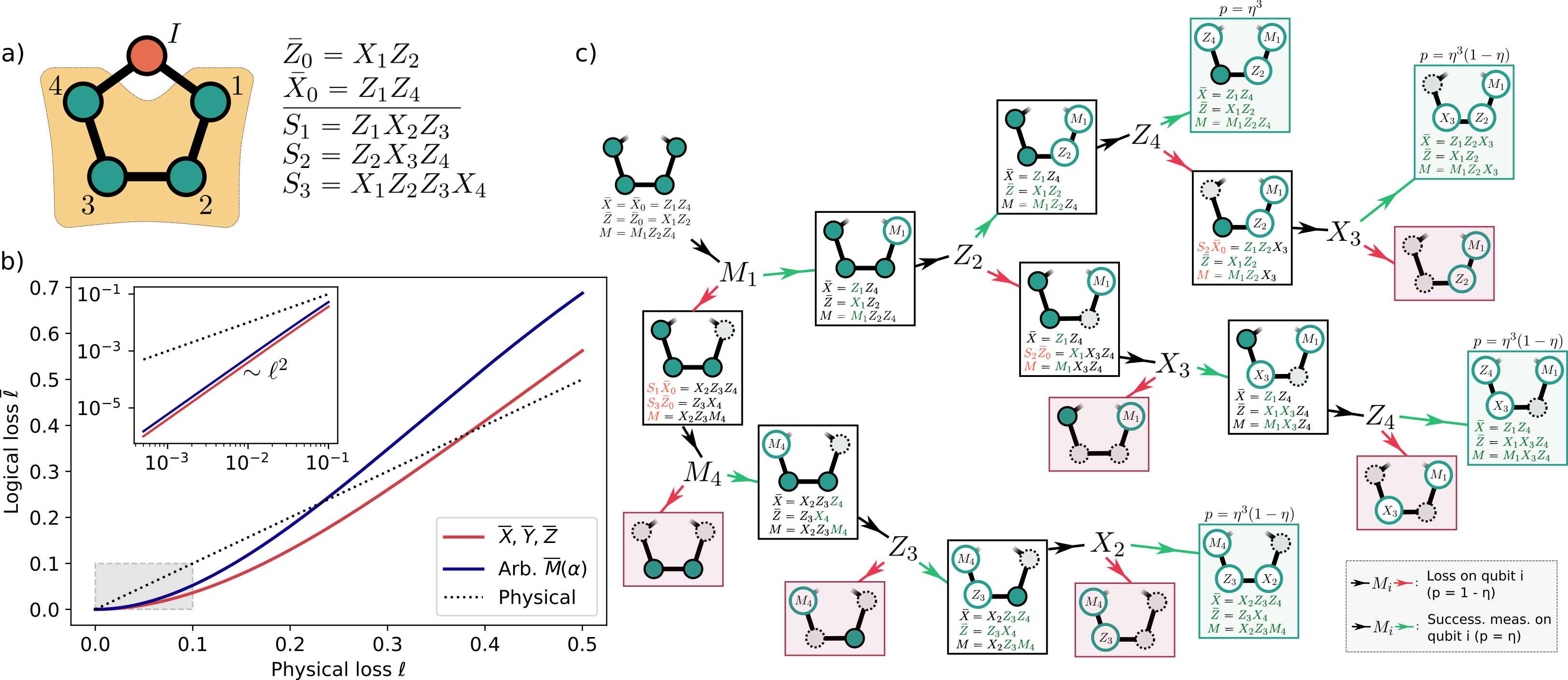}
    \caption{
    \textbf{Loss-tolerant logical measurements with the pentagon graph.} 
    a) Progenitor graph for the pentagon code, with its two logical operators and the list of stabilizer generators. 
    b) Performance of the pentagon code for all logical Pauli measurements (red) and logical measurements in an arbitrary basis (blue). 
    Physical loss is shown as a dotted black line, with break-even points obtained at $38\%$ and $23\%$ for Pauli and arbitrary measurements, respectively. 
    c) Exemplary decision tree for decoding an arbitrary measurement on the pentagon graph tolerating the loss of any single qubit.
    Green arrows represent successful qubit measurements, red arrows represent loss of the measured qubits.
    At each measurement, the number of valid measurement strategies remaining is reduced.
    }
    \label{fig::pentagon}
\end{figure*}

In arbitrary graph codes, the optimality of a measurement pattern $M$ on the remaining undetected qubits may depend on the losses detected on already-measured qubits.
Therefore, given a graph code, an important task is now finding a decoding strategy that optimises the probability to achieve a measurement pattern $M$ providing a successful logical measurement. 
Specifically, while decoding in presence of unheralded loss we need to consider: 1) a single-qubit measurement pattern $M$ consistent with the measurements already performed, 2) a rule that determines which of the unmeasured qubits should be measured next, 3) a rule that allows us to update the decoding strategy as new qubit loss is detected.
The general structure for the algorithm we consider to optimise loss-tolerant decoding strategies is described in Algorithm~\ref{alg:Paulidecoder}.
It is an iterative algorithm where at each iteration a new measurement is decided via a  \texttt{NextMeas} function, and the available operators $\overline{\mathcal{L}}[A]$ updated via \texttt{UpdateDecoder}, which implements the constraints in Eq.~\ref{eq:MBQEC_constraint_logop}.
For moderate-size codes we can analyse the decoding procedure in terms of a decision tree describing the evolution of the decoder status conditional on the qubit measurements.
The resource-intensive process of building the decision tree can be done offline prior to execution, such that runtime costs are reduced to up to $n$ queries of a look-up table, each possibly followed by adjustment of measurement bases.
Such structures also provide us an analytical formula for the logical success probability in terms of the qubit loss $\ell$ by summing the conditional probabilities of all paths in the tree that end in a successful logical measurement, as exemplified in Fig.~\href{fig::pentagon}{\ref{fig::pentagon}c} for the pentagon code.  
More details on the algorithm and decoding procedures are described in Appendix~\ref{sec:decoder}.

\begin{figure*}
    \centering
    \includegraphics[  width=1 \textwidth]{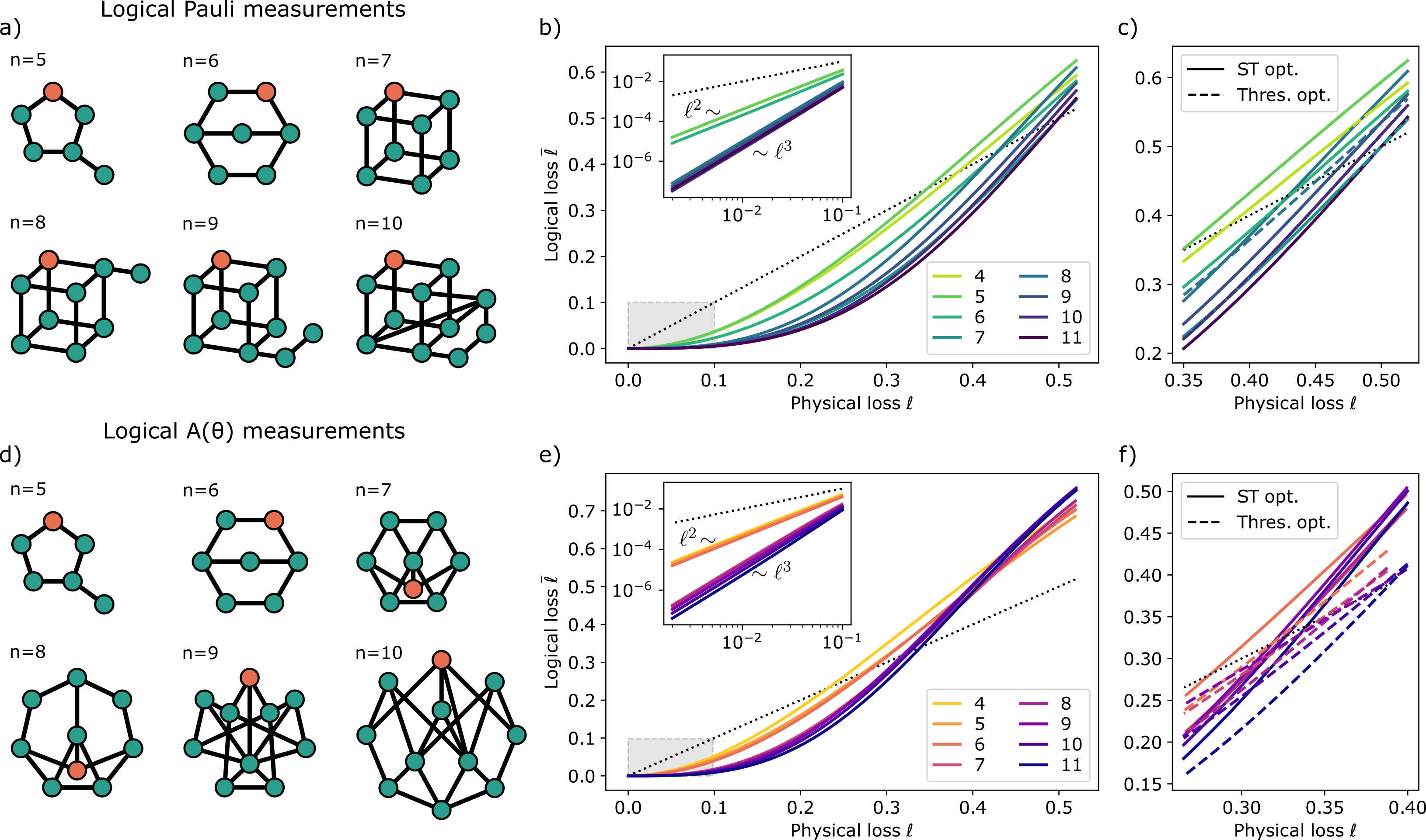}
    \caption{
    \textbf{Optimised graph codes for loss tolerant measurements.}  
    a) Progenitor graphs of the graph codes optimised for Pauli basis measurements in the sub-threshold regime for different code sizes, with the associated performance shown in b). 
    Each graph is a representative with minimum edge number in its local-equivalence class. 
    c) Comparison of near-threshold performance for graphs optimised at near-threshold loss (dashed lines) and the subthreshold-optimised graphs (solid lines). 
    d)-f) Analogous plots for arbitrary basis measurements.}
    \label{fig::optimalLT}
\end{figure*}

When testing the graphs in Fig.~\ref{fig:Paulimeas} we retrieve the expected loss tolerance with scaling $\overline{\ell} = \ell^n$, optimal for a single logical Pauli measurement.
A more interesting problem is instead to find codes with good loss tolerance for any logical Pauli $\overline{A}\in\{\overline{X}, \overline{Y}, \overline{Z} \}$ measurement.
The fact that loss tolerance is invariant in locally-equivalent graphs~\cite{Hein2004PRA,VanDenNest2004} (see Appendix \ref{sec:LUE}) allows us to make the optimisation procedure more efficient, as we only have to analyse one representative graph state per local-equivalence class for a comprehensive analysis.
Detailed categorisation of graph state equivalence classes have been performed for graphs with up to 12 qubits \cite{Danielsen2006, Adcock2020mappinggraphstate}, which we use to carry out an exhaustive search of small-scale (up to $n=11$ code qubits, i.e. progenitor graphs with 12 qubits) graph codes.
We find that 12 qubit progenitor graph states can be analysed typically in a few seconds on a standard laptop, but due to the large number of equivalence classes ($>10^{6}$) we take advantage of the high performance computing cluster BlueCrystal at the University of Bristol.
We identified the pentagon code, shown in Fig.~\href{fig::pentagon}{\ref{fig::pentagon}a}, as the smallest code ($n=4$) showing loss tolerance simultaneously for more than one logical Pauli measurement. 
For this graph we obtain the same logical success probability $\overline{\eta}= 2\eta^2 - \eta^4$ for logical measurements of any Pauli operator, plotted in Fig.~\href{fig::pentagon}{\ref{fig::pentagon}b}, where $\eta = 1-\ell$ is the physical transmittivity and $\overline{\eta} = 1-\overline{\ell}$.
When the physical loss is below $\ell^* \simeq 38\%$ loss tolerance begins to appear as the logical loss is lower then the physical one.
This value is denoted the \textit{break-even point}, where the graph encoding outperforms the bare physical qubit, and is not necessarily equal to the code's loss \textit{threshold} under concatenation, see Section~\ref{sec:concatenated}.
At low loss rates $\overline{\ell} \sim 4\ell^2$ (see Fig.~\href{fig::pentagon}{\ref{fig::pentagon}b} inset), indicating the code is able to protect against the loss of any single qubit. 
We show in Fig.~\href{fig::optimalLT}{\ref{fig::optimalLT}a-c} the results of the optimisation for progenitor graphs with up to 12 qubits.
In Fig.~\href{fig::optimalLT}{\ref{fig::optimalLT}a} we report some of the graph states (see Appendix~\ref{sec::appendix_library} for a complete list) we identified with optimised loss-tolerance at loss values smaller than the break-even point, analogous to a subthreshold regime (i.e. optimising at a physical loss level $\ell = 1\%$).
The associated performances are plotted in Fig.~\href{fig::optimalLT}{\ref{fig::optimalLT}b}.
Break-even points up to $50\%$ can already be achieved for these graph codes of moderate size.
In some cases, further improvements can also be obtained considering graphs optimised at loss levels close to the break-even point (i.e. $\ell \simeq 30\%$, depending on the code size), represented by the dashed lines in Fig.~\href{fig::optimalLT}{\ref{fig::optimalLT}c}, with the associated graph states reported in Appendix~\ref{sec::appendix_library}.

\subsection{Logical measurements in an arbitrary basis}

\label{sec:arbitrary}

In the measurement-based framework, Pauli measurements correspond to Clifford operations.
In order to perform universal quantum computations, we also require measurements in an arbitrary basis on the Bloch sphere equator $A(\theta)= X \cos(\theta) + Y \sin(\theta)$ providing non-Clifford operations~\cite{Raussendorf2001onewayqc}.
In order to perform this operation loss tolerantly on an graph code, the idea we consider can be regarded as adaptive teleportation of the encoded state into a single code qubit preemptively measured in $A(\theta)$.
If we know from the start that a physical qubit is not lost, which we call the \textit{output} qubit, then a sufficient condition for measuring $\overline{A}(\theta)$ on $\ket{\overline{\psi}}$ is to measure the other code qubits such that the encoded state is teleported on to the output qubit~\cite{morley2019}.
However, we can also think of inverting the order of these operations: first attempt to measure $A(\theta)_{\text{out}}$ and then do the teleportation only if the output photon is successfully detected, and otherwise try again with a different output qubit~\cite{Varnava2006}.
The two orderings provide the same outcome up to a feed-forward operation on the output.
Care is hence required to account for the Pauli frame update imposed by the outcome of intermediate measurements. 
This may be done with classical post-processing of measurement outcomes in some cases, in other cases more sophisticated correction circuits may need to be employed, for example, those discussed in Ref.~\cite{nielsen2003, Varnava2006}. 
Procedures to perform arbitrary logical measurements can thus be obtained by adapting techniques for loss tolerant teleportation in graph states to the case where the output is not a fixed qubit.
In particular, Ref.~\cite{morley2019} provides a sufficient condition for a measurement pattern $M$ to teleport the encoded state to a fixed output qubit, called the \textit{stabilizer pathfinding conditions} (SPC).
In terms of the constraints in Eq.~\ref{eq:MBQEC_constraint_logop}, SPC can be stated as follows: a measurement pattern $M$ of local Pauli operators on code qubits not including the output ($M_{\text{out}} = \mathds{1}$) teleports the encoded state to the output qubit, up to a random but known local unitary $U_{\text{out}} \in \{ \mathds{1}, X, Y, Z\}$, if $\overline{\mathcal{L}}_M$ contains two anticommuting logical operators. 
For logical measurements, to this condition we need to add the successful initial measurement of the output in $A(\theta)_{\text{out}}$.
The SPC can be included in the decoder very similarly as in the Pauli measurement decoder discussed in the previous section (Algorithm~\ref{alg:Paulidecoder}).
The main differences are: 1) we now need to consider the set of all logical operators $\overline{\mathcal{L}}$, instead of a single logical Pauli, and 2) we require it to contain at least two anticommuting operators at the end, instead of just being nonempty.
The approach is thus modified to a decoder structure as described in Algorithm~\ref{alg:arbitrarydecoder}, with more details reported in Appendix~\ref{sec:decoder}.

\begin{algorithm}[h!]
    \caption{Arbitrary measurement decoder}\label{alg:arbitrarydecoder}
\KwIn{Set $\overline{\mathcal{L}}$ of all logical operators, operator $A$.}
\KwOut{Outcome of arbitrary logical measurement$\overline{A}$ if successful, \texttt{False} otherwise.}

Initialise $\Theta=\{1,\ldots,n\}$, $M=\mathds{1}^{\otimes n}$, $outdet = \texttt{False}$\;

\While{$\Theta$ and $\overline{\mathcal{L}}$ are non-empty}{
    \tcc{Measure an output qubit}
    \While{!$outdet$}{
        $out \leftarrow \texttt{NextOut}(\Theta, \overline{\mathcal{L}})$ \Comment*[r]{Decide next output qubit $out$ to attempt measuring in $A_{out}$.}
        
        Attempt $A_{out}$ on qubit $out$, set $outdet = \texttt{True}$ if successful, $M_{out}= \nexists$ if qubit is lost\;  
        
        $\overline{\mathcal{L}} \leftarrow \texttt{UpdateDecoder}(M, \overline{\mathcal{L}}, out)$\Comment*[r]{Update $\overline{\mathcal{L}}$ according to Eq.\ref{eq:MBQEC_constraint_logop}.}
        
        Remove qubit $i$ from $\Theta$.
    }

    \tcc{Teleport to the output}

    $i, P_i \leftarrow \texttt{NextMeas}(\Theta, \overline{\mathcal{L}}, out)$. \Comment*[r]{Decide next qubit $i$ and Pauli operator $P_i$ to measure.}
    
    Measure qubit $i$, set $M_i= P_i$ if successful, $M_i= \nexists$ if qubit is lost\;  
    
    $\overline{\mathcal{L}} \leftarrow \texttt{UpdateDecoder}(M, \overline{\mathcal{L}}, out)$\Comment*[r]{Update $\overline{\mathcal{L}}$ according to Eq.\ref{eq:MBQEC_constraint_logop}. Set \textit{outdet} = \texttt{False} if current output no longer feasible.}
    
    Remove qubit $i$ from $\Theta$.
}
If $outdet$ is \texttt{True} and $\overline{\mathcal{L}}$ contains a pair of anticommuting operators return value of $A_{out}$ and $U_{out}$ from the SPC, else return \texttt{False}.

\end{algorithm}
%

The loss-tolerance of graph codes under arbitrary basis measurements is again preserved between locally-equivalent graphs (see Appendix~\ref{sec:LUE}), allowing for a streamlined optimisation procedure.
The smallest code displaying loss tolerance we identify is again the pentagon progenitor graph in Fig.~\href{fig::pentagon}{\ref{fig::pentagon}a}.
The logical success probability for arbitrary $\overline{A}$ measurement with this graph is $\overline{\eta} = 4 \eta^3 - 3 \eta^4$ (see Fig.~\href{fig::pentagon}{\ref{fig::pentagon}b}).
Loss tolerance is observed below a physical loss breakeven point of $\ell^* \approx 23\%$, and we observe a subthreshold scaling of the logical loss as $\sim 6 \ell^2$ indicating tolerance against the loss of any single code qubit.
It can be noted, also comparing the behaviours in Fig.~\href{fig::pentagon}{\ref{fig::pentagon}b}, that the loss tolerance performance is worse than logical measurements of Pauli operators, as expected.
In Fig.~\href{fig::pentagon}{\ref{fig::pentagon}d} we report, for various code sizes, the graphs we identified which optimise the loss tolerance for arbitrary logical measurements in the subthreshold regime. 
Their logical loss behaviour is shown in Fig.~\href{fig::pentagon}{\ref{fig::pentagon}e}.
We again observe higher loss tolerance for larger codes, and thresholds higher by a few percent when optimising in a loss range close to break-even points, as shown in Fig.~\href{fig::pentagon}{\ref{fig::pentagon}f}.  
For the largest size explored, $n=11$ code qubits (12 qubits progenitor graphs), we obtain a break-even point of $\ell^* \simeq 40\%$.

\section{Measurement-based error correction in loss tolerant graphs}
\label{sec::errors}

\begin{figure*}
    \centering
    \includegraphics[  width=1 \textwidth]{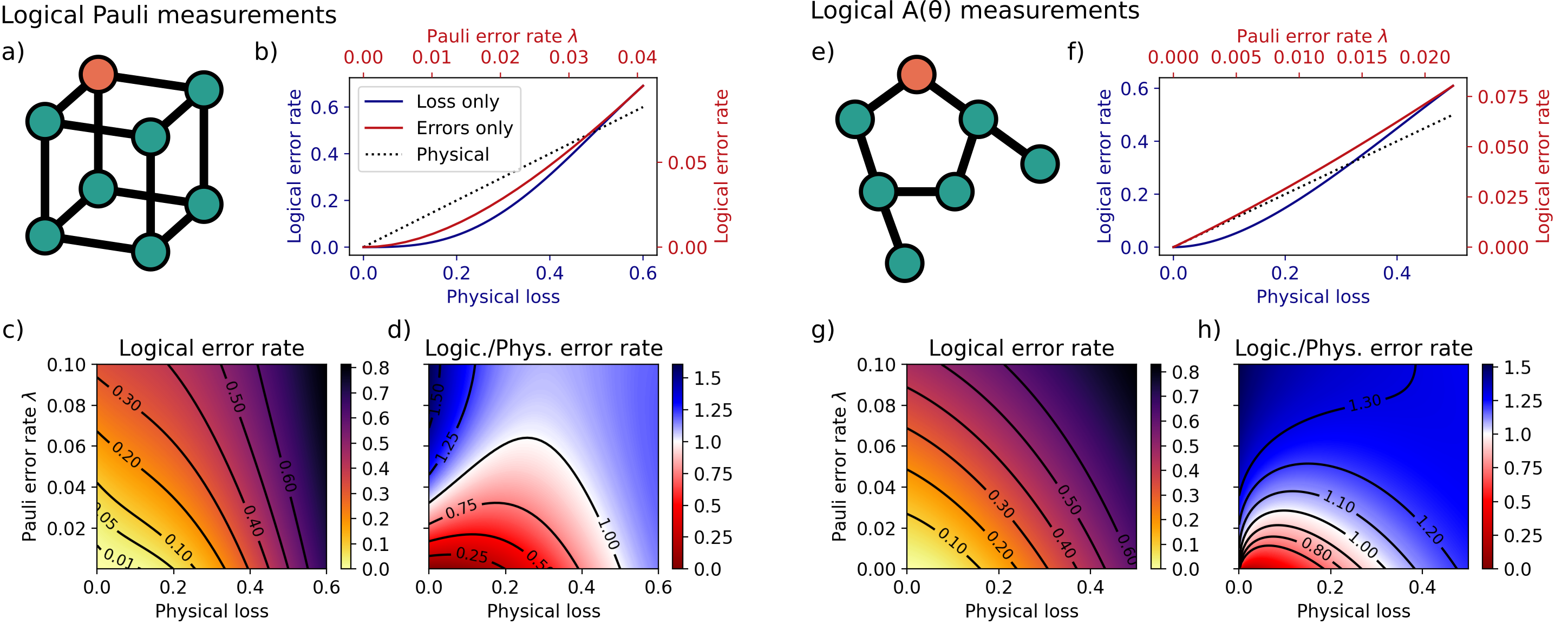}
    \caption{\textbf{ Simultaneous correction of loss and errors on graph codes.} 
    a) The cube graph the smallest progenitor graph able to perform logical measurements in all Pauli bases correcting simultaneously losses and errors. 
    b) Logical error probability for each noise type in isolation,
    and c) when both are simultaneously present.
    d) Ratio of encoded to bare error rates, highlighting the region (in red) where error suppression occurs. 
    e) - h) Analogous plots for arbitrary basis measurements.
    The smallest graph able to correct losses and to saturate the scaling $\overline{\epsilon}/\epsilon\rightarrow1$ at low error rates is the decorated pentagon graph.}
    \label{fig::errors}
\end{figure*}

Qubit errors arising from imperfect gates and measurements can also be simultaneously corrected with graph codes.
Unlike photon loss, they cannot be directly detected, so need to be inferred using the code stabilizers $\mathcal{S}_M$ and operators $\overline{\mathcal{L}}_M$ induced by the measurement pattern $M$, as described in Section~\ref{sec:background}.
For logical Pauli measurements, measurement-based error correction corresponds to updating the Pauli frame, meaning that the correction can be done by post-processing the logical measurement outcome. 
Specifically, after inferring an error $E$ from a decoder, the outcome of $\overline{L}$ is flipped if the supports of $E$ and $\overline{L}$ share an odd number of qubits (if no $\overline{L}$ exists due to losses, we consider it a logical error as well). 
For arbitrary logical measurements, the situation is similar but considers both logical operators $(\overline{L}, \overline{K})$ in a pair satisfying the SPC, as described in Section~\ref{sec::measurements}. 
Note, however, that in this case we cannot identify and correct errors on the physical $A_\text{out}(\theta)$ measurement of the output qubit,  as that operation is outside the stabilized space (unless $A_\text{out}(\theta)$ is a Pauli measurement).
Therefore, the logical error rate for arbitrary measurements cannot be smaller than the physical rate on the output qubit.
This is related to the fact that in the measurement-based framework logical arbitrary measurements $A_\text{out}(\theta)$ correspond to arbitrary non-Clifford single-qubit operations, and stabilizer codes possess only a limited set of natively fault-tolerant gates~\cite{EastinKnill, BravyiKonig}.
To maintain generality for arbitrary graph codes, we implement error decoding via maximum-likelihood, which is computationally viable for the moderate-size codes considered here, and consider a phenomenological error model of i.i.d. Pauli errors on each qubit, corresponding to the depolarising channel
$\rho\rightarrow \left(1-3\lambda\right) \rho + \lambda\left(X\rho X + Y\rho Y + Z\rho Z\right)$.
More details on the decoding procedure can be found in Appendix~\ref{sec:decoder}.
The error correction is again invariant for locally equivalent graphs, up to permutations of the Pauli bases which arise from local complementations on the progenitor graph,  which effectively act as Clifford operations on the logical operators (see Appendix~\ref{sec:LUE}).
We can thus analyse individual graphs from local-equivalence classes to characterise codes of increasing size, now using models where errors and loss are simultaneously present.
Considering indirect Pauli measurements, the smallest progenitor graph found to exhibit fault tolerance against both errors and losses is the cube graph, shown in Fig.~\href{fig::errors}{\ref{fig::errors}a}, which generates a code locally equivalent to the seven-qubit Steane code~\cite{Steane96}.
As shown in ~\href{fig::errors}{\ref{fig::errors}b}, when noises are individually present, it outperforms the bare qubit for losses below $50\%$, saturating the bound of the measurement complementarity principle~\cite{nickerson2018}, and for physical errors $\lambda\leq3.2\%$.
In Fig.~\href{fig::errors}{\ref{fig::errors}c} we plot the overall fault probability in the presence of loss and Pauli errors, where the fault probability is the probability of at least one error type occurring during measurement.
In Fig.~\href{fig::errors}{\ref{fig::errors}d} we show the ratio between the logical and physical fault probabilities, where now a break-even curve can be observed and shows a remarkable robustness for this code.
As mentioned above, it is not possible to reduce logical errors below the single qubit level for arbitrary basis measurements.
Still, we find examples of graphs that saturate this linear bound at low error rates.
For example, we show in Fig.~\href{fig::errors}{\ref{fig::errors}e} the decorated pentagon graph, which is loss-tolerant for arbitrary measurements with a break-even point of 32$\%$ (see Fig.~\href{fig::errors}{\ref{fig::errors}f}) while simultaneously having logical error rates $\overline{\epsilon}/\epsilon \rightarrow 1$ as $\epsilon\rightarrow 0$.

\section{Extending to larger graphs by modularisation}
\label{sec:concatenated}

\begin{figure*}[]
    \centering
    \includegraphics[width = 1\textwidth]{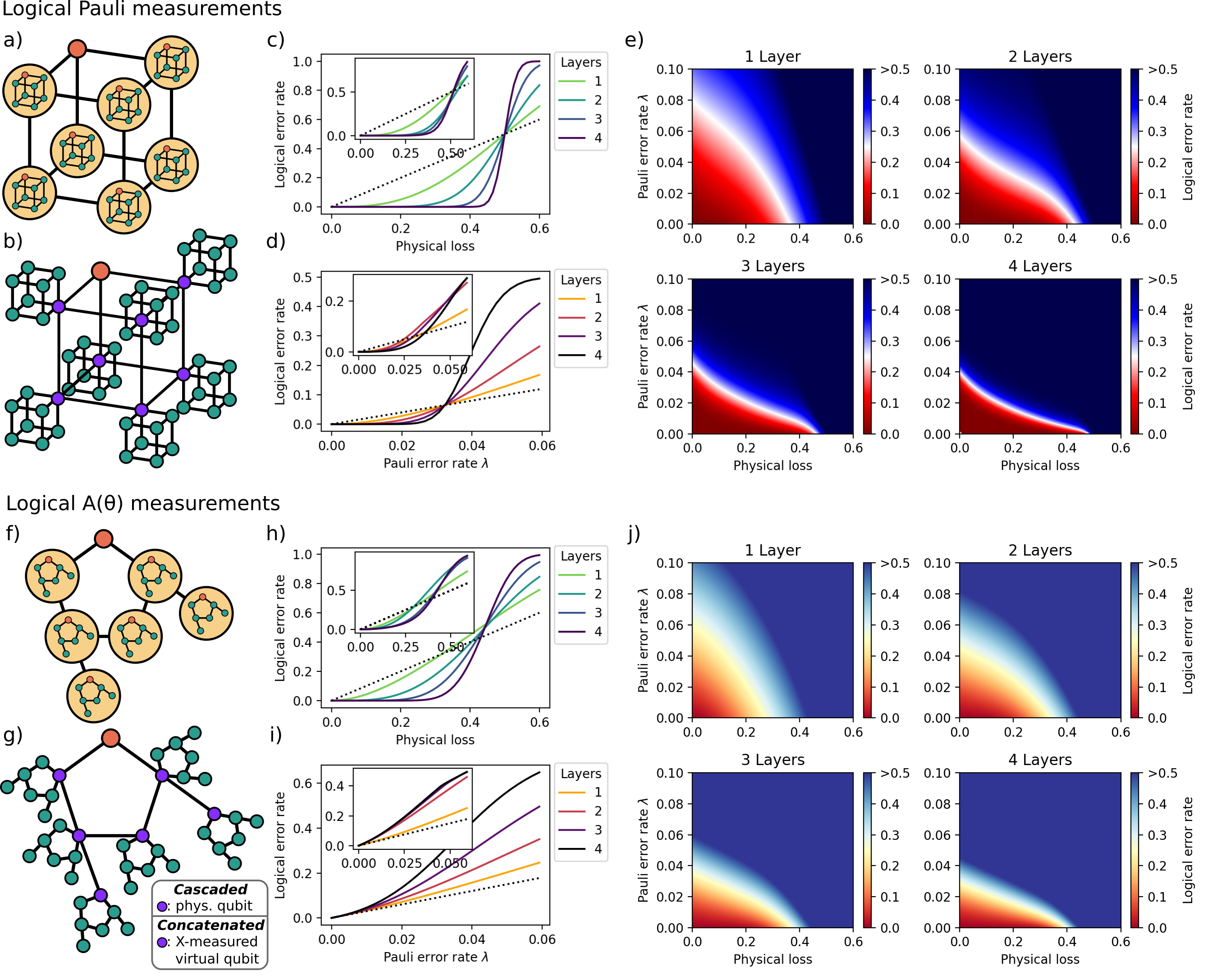}

    \caption{\textbf{Performance of cascaded and concatenated codes.}
    a) Modular construction using the  cube graph code, and b) the resulting progenitor graph state with two layers.
    The qubits in layer 1, shown in purple, are physical for the cascaded construction, and virtual X-measured qubits for the concatenated one.
    c) Loss-tolerance performance, considering a loss-only noise model, for logical Pauli measurements self-concatenating the cube graph code at different layers, showing a threshold at $50\%$ loss.
    The performance for equivalent cascaded constructions is shown in the inset.
    d) Performance for error-tolerance, with an i.i.d. error-only model, showing a threshold at $\lambda \simeq 3.2\%$
    e) Overall fault probability in the presence of both noise types for concatenations of up to depth 4, showing the emergence of a threshold curve. 
    e)-j) Equivalent plots for the decorated pentagon code for arbitrary basis measurements. 
    }
    \label{fig::cascaded}
\end{figure*}

To analyse the performance of larger codes efficiently, we consider two modular approaches: cascading and concatenating small \textit{unit graphs} which can be fully characterised using the previously described techniques.

\subsection{Cascaded graphs}

We define cascaded graphs as layered graphs constructed by recursively appending unit graphs to code qubits.
In particular, by embedding each code qubit in the $k$-th layer with a unit graph to which it is the input qubit, then the code qubits of all the added unit graphs represent the qubits in the $(k+1)$-th layer.
This step can then be repeated to recursively build larger graphs in a modular approach, where the unit graphs used can also vary at different layers.
Examples of cascaded graphs are shown in Fig.~\href{fig::cascaded}{\ref{fig::cascaded}b, g}.
The inspiration for the cascaded graphs construction comes from tree graph codes~\cite{Varnava2006}, which can be seen as cascades of star-graphs (see Fig.~\href{fig:Paulimeas}{\ref{fig:Paulimeas}a}), locally equivalent to Greenberger-Horne-Zeilinger (GHZ) states.
Cascaded graphs are a generalisation of tree graph structures using arbitrary graphs as modules.
The structure of cascaded graphs is such that the analysis of their loss and error tolerance properties can be efficiently obtained once the performance of the small-size unit graphs is known.
The idea is to consider the measurement patterns suggested on the top-layer graph, and modify them recursively when going to deeper layers.
In fact, the cascaded structure allows us to leverage indirect measurements of qubits in upper layers via the measurement of qubits in deeper layers.
For example, if a top-layer qubit is to be measured in the $Z$ basis, it can equally be measured indirectly using qubits restricted to the second layer of the graph, so loss of the qubit can be tolerated.
In fact, for any lost qubit at depth $k$ of a cascaded structure, one can attempt to recover an indirect $Z$ measurement by measuring depth $k+1$ qubits. 
Measurements in non-$Z$ bases, however, necessarily require additional indirect 
measurements of deeper qubits, as logical non-$Z$ operators in the progenitor graph have always support on the code qubits (the set $B$ in Fig.~\href{fig:graph_codes}{\ref{fig:graph_codes}c}).
Such measurements are effectively disentangling deeper layers, analogously as in tree graphs~\cite{Varnava2006}.
The asymmetry between $Z$ and non-$Z$ bases can thus be attributed to the geometry of the stabilizers in the full graph; $X$- and $Y$-type stabilizers penetrate between layers, whereas $Z$-type stabilizers do not.
Using the properties described above, we can perform the decoding process for cascaded graphs recursively using only the properties of unit graphs at different layers in the cascade. 
We report such recursive functions, for decoding both losses and errors, in Appendix~\ref{sec::appendix_cascade}.
In Fig.~\href{fig::cascaded}{\ref{fig::cascaded}c, d} (insets) we report an example of improved logical losses and error rates obtained by cascaded graph structures. 
Here the unit graph is taken to be the smallest graph we identified in Section~\ref{sec::errors} with tolerance to both loss and errors for logical Pauli measurements - the cube progenitor graph.

\subsection{Concatenated graphs}

\begin{figure}
    \centering
    \includegraphics[width=1 \columnwidth]{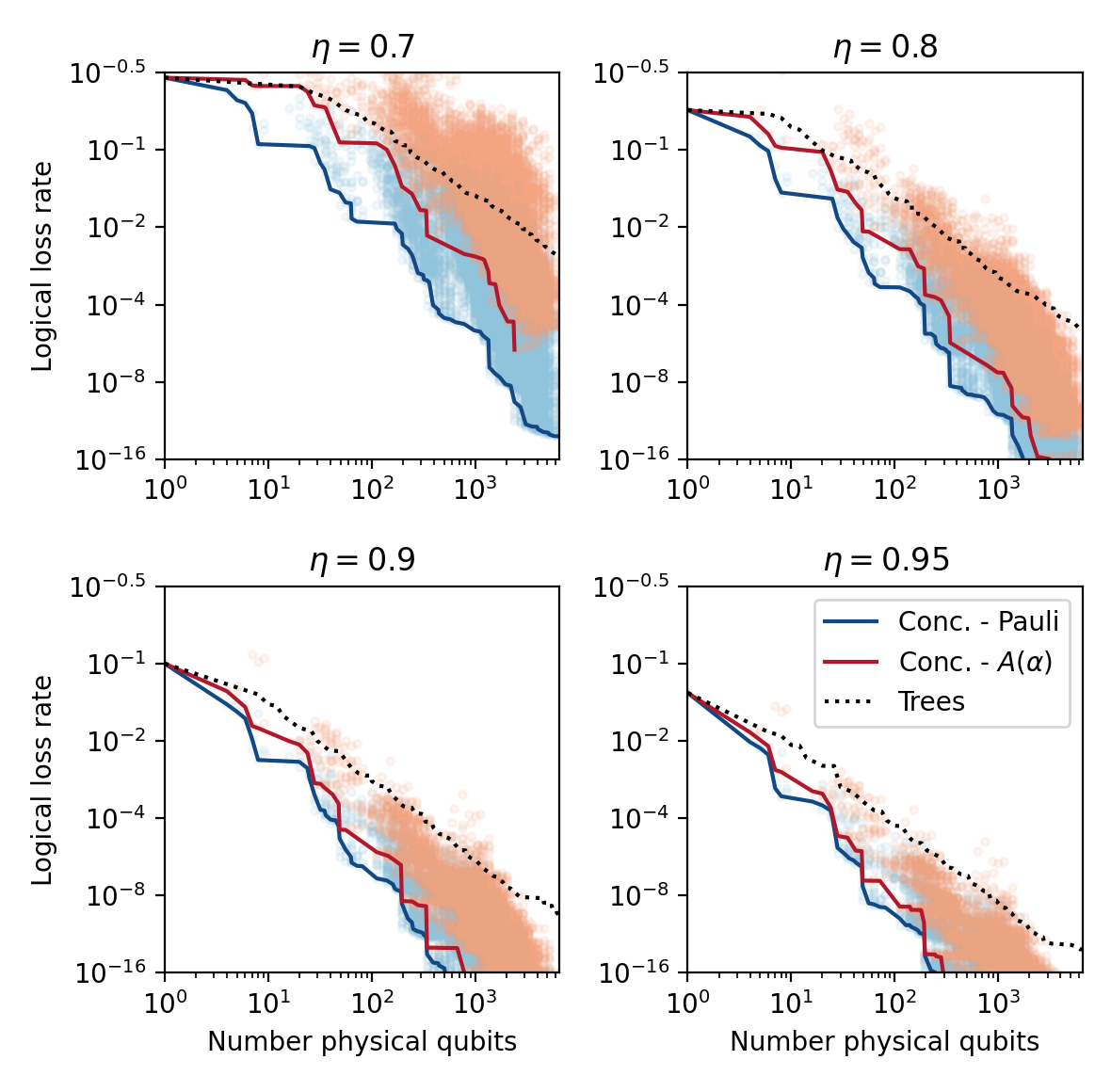}
    \caption{
    \textbf{Resource scaling} for optimised concatenated graph codes for logical Pauli (blue) and arbitrary $A(\alpha)$ (red) measurements, compared with the performance of optimised tree graphs (dashed black) from Ref.~\cite{Varnava_2007}. 
    Semi-transparent points correspond to logical losses for all the combinations of concatenated graphs tested in the optimisation procedure, and solid lines correspond to the optimised scaling from the best concatenation configurations obtained.
    The numerical scaling data, which considers exclusively losses, is reported for different values of the per-photon transmission $\eta$. 
    Note the scale for all plots is log versus loglog.
    }
    \label{fig:trees_vs_concat}
\end{figure}

Modular extensions of graph codes can also be performed via graph code concatenation where each code qubit is itself encoded in another code -- a standard approach in QEC.
Concatenation of graph codes can be described by simple graph operations and be used for constructing concatenated quantum codes of increasing size~\cite{Beigi2011}.
A concatenated graph code can be also easily described starting from the cascaded construction of the previous section: it corresponds to considering every qubit in intermediate layers as virtual qubits measured in the $X$ basis and with $+1$ outcome obtained.
The measurement-based decoding procedures can also be performed similarly in a recursive fashion, with the only difference that now all measurements performed on intermediate layers are indirect and direct measurements are only performed on the qubits in the deepest layer (see Appendix~\ref{sec::appendix_concatenation} for details).
Note in fact that virtual qubits do not have to exist in practice and are only useful in describing the concatenated graph; only the lowest depth qubits are physical. 
We show examples of concatenated codes in  an encoding is shown in Fig.~\href{fig::cascaded}{\ref{fig::cascaded}b, g} for self-concatenations of the cube (Steane) and decorated pentagon graph codes.
In Fig.~\href{fig::cascaded}{\ref{fig::cascaded}c-e} we report the performance for concatenated cube graph codes in the presence of losses and errors, for up to 4 layers of concatenation when performing Pauli-basis measurements.
We observe a threshold appearing for losses at 50$\%$, saturating the bound set by the measurement complementarity principle (see Appendix~\ref{sec:complementarity}), and an error threshold of $\lambda=3.2\%$.
Figures~\href{fig::cascaded}{\ref{fig::cascaded}h-j} show equivalent plots but for arbitrary logical measurements concatenating the decorated pentagon graph.
As the decoding procedure described above can easily incorporate the concatenation of different unit graphs at different layers, we can optimise the combination of unit graphs to obtain higher noise tolerance at various number of code qubits. 
In Fig.~\ref{fig:trees_vs_concat} we plot the results of such optimisation for concatenated graphs at different loss levels, for both logical Pauli and arbitrary measurements.  
The optimisation was performed by directly testing all combinations for up to 4 layers of unit graphs from the set of all loss-tolerant graphs we identified from the analysis in Section~\ref{sec::measurements}.
For comparison, we also report the performance of optimised tree graphs from Ref.~\cite{Varnava2006}.
Already in the regime with few tens of qubits, we see orders of magnitude improvement in the logical loss of the optimised concatenated graphs against tree graphs for the tested physical transmission rates in the  0.7-0.95 range.

\section{Logical graph state fusions}
\label{sec:fusions}

The techniques described in previous sections for single-qubit logical measurements can also be used to analyse another key operation for photonic measurement-based approaches: \textit{fusion gates}~\cite{Browne2005,GimenoSegovia2015,fbqc}.
Introduced in Ref.~\cite{Browne2005}, they are probabilistic two-qubit entangling gates that can be implemented using simple linear-optical circuits, and have the effect of joining two photonic graph states with a destructive measurement on two photons, one from each graph.
Standard fusion operations, implementable with simple linear circuits and no ancillary resources, succeed with a probability of 50\%~\cite{Browne2005}, which can be boosted by using ancillary photons.
In particular, boosted fusions require $2^{m}-2$ additional ancillary photons and have a success probability $(1 - p_{\text{fail}})\eta^{1/p_{\text{fail}}}$, where $p_{\text{fail}}=2^{-m}$ is the probability of gate failure and $\eta$ is the transmission~\cite{Grice2011, Ewert2014, Olivo2018}.
In operator terms, a successful fusion measurement retrieves two parity measurements for the operators $XX$ and $ZZ$. 
If the gate fails, only one of the two outcomes is available, and single-photon operations can be used to choose it to be $ZZ$ or $XX$, while the other outcome is erased.
If either of the qubits is lost, the gate fails completely and neither operator is recovered.
The two mechanisms of qubit loss and gate failure  are thus inequivalent and with different consequences for the growth of clusters.
Graph codes can be used to make the fusion of the encoded logical qubits robust against both mechanisms~\cite{fbqc, hilaire2022}.
A logical fusion of two encoded qubits is a measurement providing joint parity checks $\overline{X}\overline{X}$ and $\overline{Z}\overline{Z}$.
In contrast to logical single-qubit measurements, this operation requires physical fusion gates between qubits from the two codes, i.e. physical fusion gates.
Nevertheless, as we will show, these can be readily included with the techniques developed in previous sections.
We consider two different strategies for it, as illustrated in Fig.~\href{fig::logical_fusions}{\ref{fig::logical_fusions}a} and described below.

\subsection{Transversal physical fusions}
Considering two identical graphs encoding the logical qubits to be fused, we first consider a ballistic method where physical fusions are attempted transversally between all code qubits in one graph and the equivalent qubits in the other, as shown in Fig.~\href{fig::logical_fusions}{\ref{fig::logical_fusions}a}.
Each physical fusion can be successful, fail, or be erased due to loss of one of the photons with respective probabilities $\eta^{1/p_{\text{fail}}}(1-p_{\text{fail}})$, $\eta^{1/p_{\text{fail}}}p_{\text{fail}}$, and $1-\eta^{1/p_{\text{fail}}}$.
Once all transversal fusions are performed, the logical fusion is successful if the obtained operators can generate both $\overline{X}\overline{X}$ and $\overline{Z}\overline{Z}$, fails if only one of them can be generated, or is completely lost if neither of them can.
We numerically calculate the probability for each of these three logical outcomes by considering all possible combinations of the three outcomes from fusing all $n$ pairs of code qubits in the graphs. 
The total logical fusion success probability is obtained by summing the probability associated to all combinations that lead to a successful fusion, and similarly for the logical failure and logical loss probability.
We allow the measurement recovered on physical fusion failure to be chosen independently for each qubit, which is pre-compiled using maximum likelihood before runtime to maximise the probability of successful fusion.
We report in  Fig.~\href{fig::logical_fusions}{\ref{fig::logical_fusions}b} results for logical transversal fusion on the graphs optimised for arbitrary single-qubit measurements up to $n=9$ (see Fig.~\href{fig::optimalLT}{\ref{fig::optimalLT}d}). 
Despite the approach being non-adaptive, these small codes present success probabilities that significantly outperform typical boosted fusion schemes, which we also report in the black dashed curve, in terms of loss-tolerance.
For example, using standard physical fusions with 50\% success probability, we obtain logical fusion measurements with success probability $\overline{p}_{\text{succ}}>0.8$ for physical loss $\ell=5\%$ and $\overline{p}_{\text{succ}}>0.95$  at $\ell=1\%$.

\subsection{Adaptive physical fusions}

A second approach we investigate for logical fusions is an adaptive strategy based on the ideas introduced for performing logical arbitrary basis measurements in Section~\ref{sec:arbitrary}.
Recall that the SPC identifies pairs of logical operators that anticommute on a single qubit~\cite{morley2019}, which we denoted as output qubit.
Taking two copies of a graph code, a logical fusion can be achieved by fusing an identified output qubit with the corresponding qubit in the other graph, followed by single qubit measurements on the remaining qubits.
The decoder for adaptive fusion is similar to that presented in Algorithm \ref{alg:arbitrarydecoder}: the idea, in the progenitor graph picture, is to teleport the virtual input qubits of the two codes into some output qubits which have been pre-fused together.
Explicitly, a physical fusion measurement is attempted between pairs of output qubits, and once a fusion is successful single-qubit Pauli measurements are attempted sequentially on the remaining qubits of each code, effectively implementing a separate decoder as in Algorithm \ref{alg:arbitrarydecoder} for each graph. 
This approach leads to improved loss-tolerance compared to the transversal case as, when performing single-qubit measurements, the loss of either qubit in a pair does not erase the information obtainable from the other.

\begin{figure}[]
    \centering
    \includegraphics[width = 1\columnwidth]{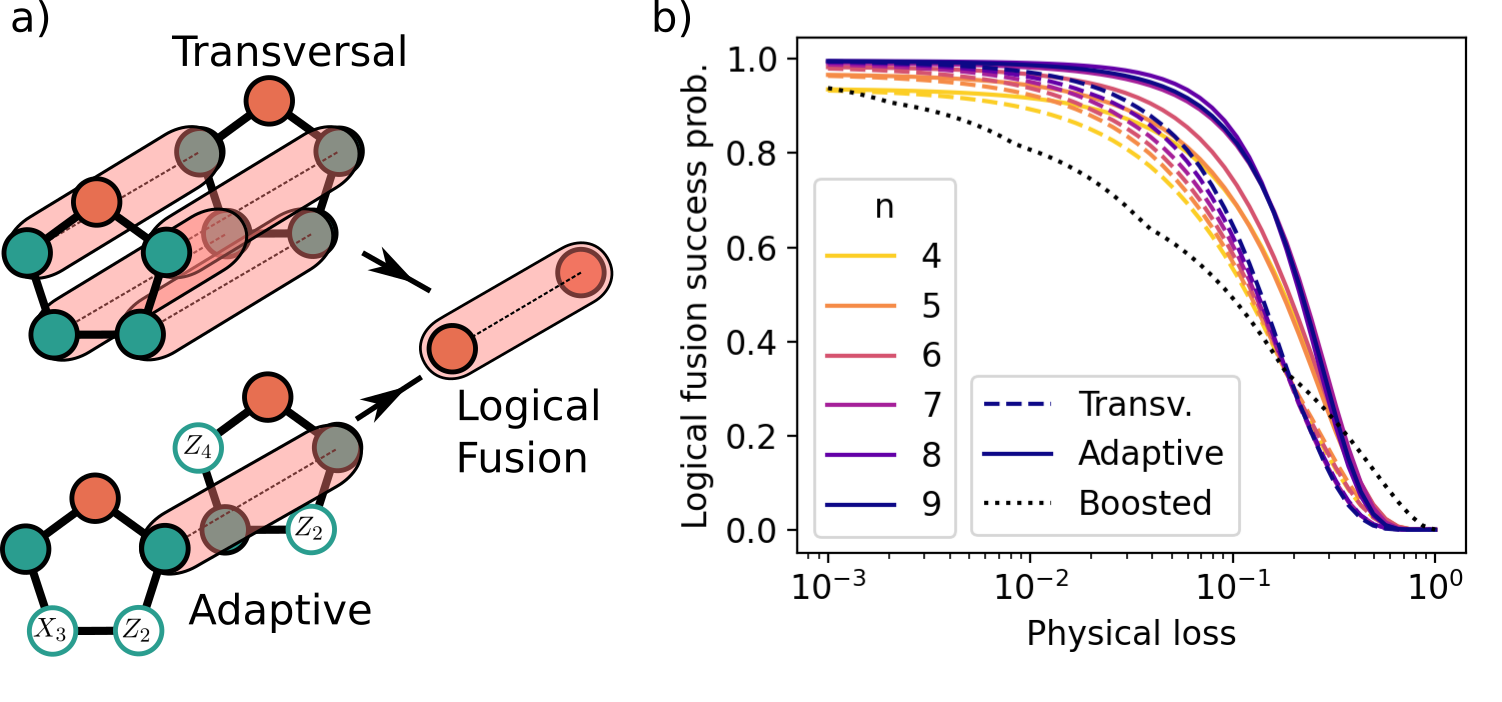}
    \caption{\textbf{Logical fusion measurements}.
    a) Two approaches to perform logical fusion operations are to perform physical fusion transversally in a ballistic manner (top), or using an adaptive strategy requiring only a single successful physical fusion and with single-qubit measurements on the remaining qubits (bottom). 
    b) The probability of logical fusion $\overline{p}_{\text{succ}}$ is plotted for the adaptive (solid lines) and transversal (dashed) approaches, for varying code sizes.
    Standard physical fusions with 50\% success probability are considered between physical qubits.
    For comparison, we also report the optimal performance of boosted fusion measurements (black dotted line) using $2^{m-1}$ ancilla qubits for $m\leq 8$.}
    \label{fig::logical_fusions}
\end{figure}

In Fig.~\ref{fig::logical_fusions} we report the performance of the adaptive strategy using the graphs optimised for arbitrary single-qubit measurements up to $n=9$ (see Fig.~\href{fig::optimalLT}{\ref{fig::optimalLT}d}).
It can be observed that the adaptive strategy generally provides better performance compared to the transversal one and boosted fusions.  
For example, using standard physical fusions with 50\% success probability, we can reach logical fusion success probabilities of $\overline{p}_{\text{succ}}=0.86$ already at a physical loss of $\ell=10\%$, and $\overline{p}_{\text{succ}}>0.99$ at $\ell=1\%$.

\section{Applications}

\label{sec::applications}

To benchmark the tools we developed for analysing and optimising general loss-tolerant graph codes, we investigate how they can be used in two exemplary applications. 

\subsection{Optimising repeater graphs}

\begin{figure}[]
    \centering
    \includegraphics[width = 1\columnwidth]{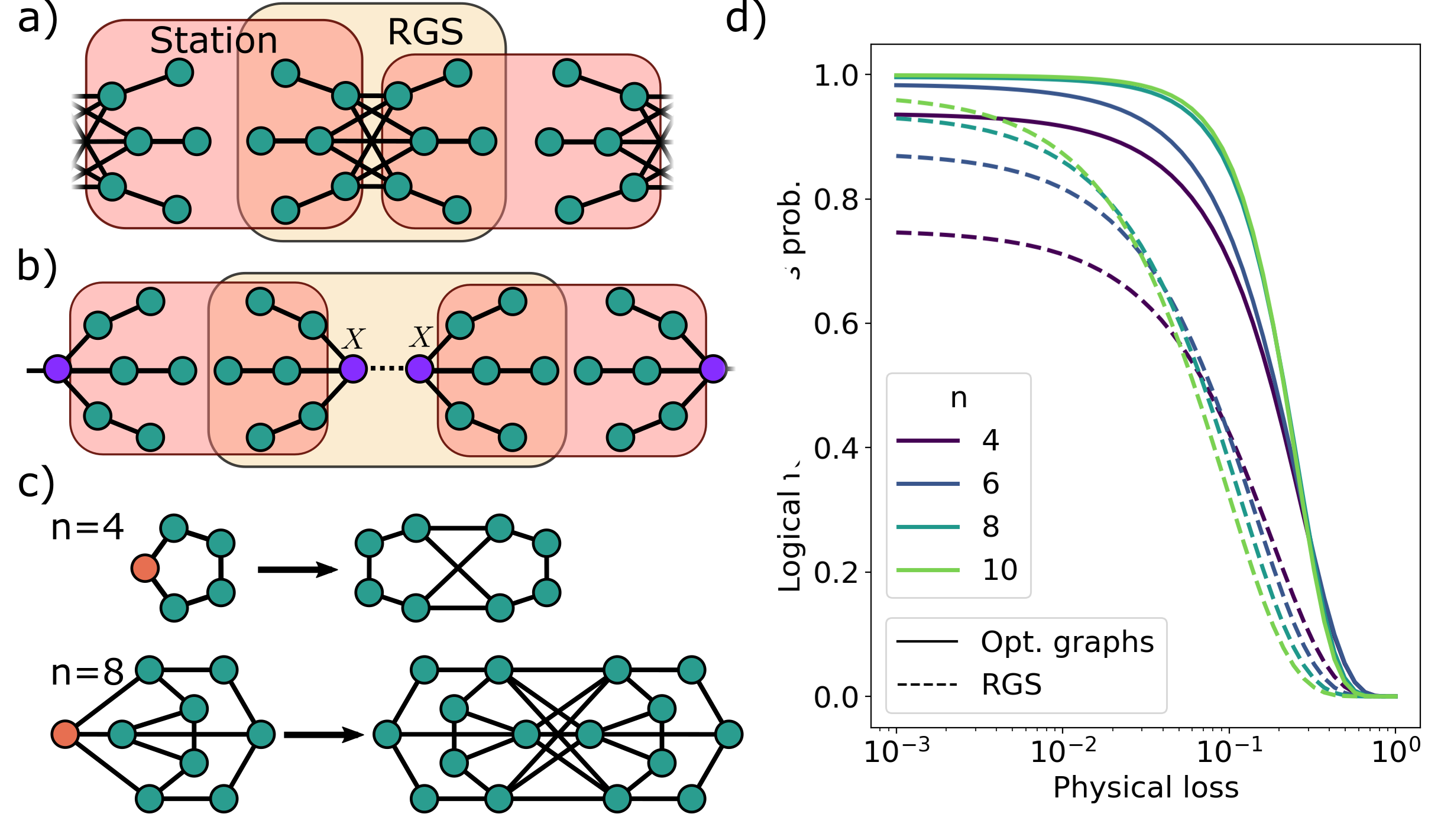}
    \caption{\textbf{Generalised repeater graph states.} 
    a) Operation of the $N=3$ repeater graph state (RGS, in the yellow box) introduced in Ref.~\cite{Azuma2015}. 
    Single qubit Pauli and fusion measurements are performed on qubits to fuse each RGS with its neighbours at consecutive repeater stations (red boxes).
    b) The progenitor graph construction of RGSs, where logical fusions are performed at each repeater station between progenitor graphs whose input qubits have been joined via a controlled-phase gate. 
    c) Generalization of RGS states constructed from loss-tolerant graph codes with $n=4$ and $n=8$ code qubits.
    d) Link success probability as a function of the physical loss rate for optimised graph codes considering adaptive non-boosted physical fusions (solid lines) against RGS constructions (dashed lines), varying number of code qubits.}
    \label{fig:rgs}
\end{figure}

Repeater graph states (RGSs) have been introduced in Ref.~\cite{Azuma2015} as an approach to making all-optical two-way quantum repeaters in a quantum network. 
The graph structure originally proposed is shown in Fig.~\href{fig:rgs}{\ref{fig:rgs}a}, and the repeater protocol works by transversely fusing the left-ward leaf (i.e. single-edged) qubits from a repeater station with the right-ward leaf qubits from the previous station.
The inner qubits are measured in $X$ if the associated leaf is the first to be successfully fused, or otherwise in $Z$ to remove unsuccessful or redundant fusions~\cite{Azuma2015}.

This protocol can also be interpreted as sequences of logical fusion operations described in Section~\ref{sec:fusions}, and can be readily analysed in the progenitor graph picture as depicted in Fig.~\href{fig:rgs}{\ref{fig:rgs}b}. 
The left-ward and right-ward qubits each correspond to physical qubits in a graph code encoding a single logical qubit.
Entanglement swapping between successive repeater stations simply corresponds to a logical fusion operation using the right-ward and left-ward codes. 
Within the repeater graph, the transmission of the logical information between the left-ward and right-ward codes can be simply described, again using the progenitor graph picture, as adding a link (i.e. a controlled-phase gate) between the input qubits of the left-ward and right-ward codes and then measuring both of them in $X$ to transmit the encoded logical qubit between left and right~\cite{Varnava2006}.
In practice, the inputs are just treated as virtual qubits, and we directly consider the total repeater graph, i.e. the graph obtained after the controlled-phase and the $X$ measurements are performed. 
The probability to successfully transmit between two consecutive stations thus simply corresponds to the logical fusion success probability of the underlying code as analysed in Section~\ref{sec:fusions}.

For the repeater graph considered in Ref.~\cite{Azuma2015}, it is easy to see that, up to local operations~\footnote{Formally, the common RGS is an $N$-qubit fully connected interior, with leaf qubits on each interior qubit~\cite{Azuma2015}. An identically-performing graph (with the exact same decoding procedure) is the graph with a `crazy graph' interior, each with a leaf qubit, which is shown in Fig.~\href{fig:rgs}{\ref{fig:rgs}a}. The progenitor graphs of each of these are the graph with $N/2 + 1$ qubit fully-connected interior (including the input node) and leaves on each code qubit, and the tree graph with branching ratios $[N/2, 1]$, respectively. These progenitor graphs are locally equivalent and thus have identical loss-tolerance performance.}, it corresponds to using tree graph codes with a branching ratio $[N/2, 1]$ as both left-ward and right-ward codes (see Fig.~\href{fig:rgs}{\ref{fig:rgs}b}), with $N$ the branching of the repeater graph~\cite{Azuma2015}.
However, as discussed in the previous sections, tree codes are suboptimal for loss tolerance, and better performance can be obtained using codes optimised for logical fusion success probability.
Using the construction presented above, we show in Fig.~\href{fig:rgs}{\ref{fig:rgs}c} the repeater graphs obtained from two optimised codes for logical fusion with $n=4$ (the pentagon graph) and $n=8$.
Their performance with adaptive fusion strategies is reported in Fig.~\href{fig:rgs}{\ref{fig:rgs}d}, in which we also show for comparison the performance of standard RGSs.
We see significant improvements in link generation probability compared to the standard tree-based RGSs with the same number of physical qubits, showing that our tools can bring significant improvements in the design of all-optical repeater schemes.

\begin{figure*}[]
    \centering
    \includegraphics[width = 1\textwidth]{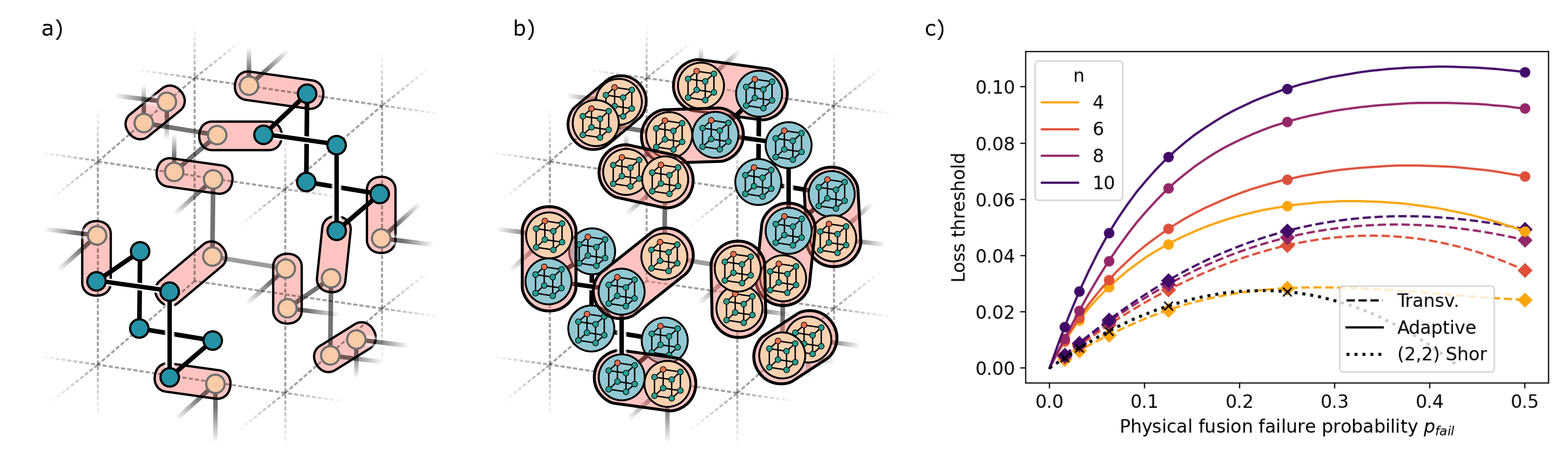}
    \caption{\textbf{Fusion-based fault-tolerance.} 
    a) Construction of a fault-tolerant fusion network for FBQC via fusing six-qubit hexagonal resource state, from Ref.~\cite{fbqc}.
    b) Concatenation of the fusion network with graph codes to enable logical fusions tolerant to loss and failure. 
    c) Per-photon loss tolerance thresholds for concatenations with optimised graph codes of varying sizes, using the adaptive decoding strategy (solid lines) and the transversal strategy (dashed). 
    The performance of the (2,2) Shor encoding presented in Ref.~\cite{fbqc} is also showed as black dotted line.}
    \label{fig:FTconcatenations}
\end{figure*}

\subsection{Fusion-based fault-tolerant schemes}
Fusion-based quantum computation (FBQC) is a variant of measurement-based quantum computing where the computation is  performed via probabilistic fusion gates between separate resource states rather than single-qubit measurements on large entangled cluster states~\cite{fbqc}.
It has been recently introduced as a convenient picture to describe photonic quantum computation as it facilitates a direct description of probabilistic fault-tolerant architectures fusing small resource states, enabling a simple treatment of failed fusion operations and qubit loss.
However, in the constructions from the original proposals~\cite{fbqc}, the per-photon loss thresholds are limited to $<1\%$, also requiring highly boosted fusions.
To improve them, concatenating qubits with a (2,2) Shor code was proposed, whereby a loss-tolerance threshold of 2.7\% per photon is achieved for boosted physical fusions with $75\%$ success probability.
To obtain better performances, we can use the techniques developed in previous sections to consider concatenating resource states with more general graph codes, as shown in Fig.~\href{fig:FTconcatenations}{\ref{fig:FTconcatenations}a,b}.
In FBQC, the $XX$ and $ZZ$ parity measurement outcomes obtained from qubit fusions are used to construct the primal and dual syndrome graphs of a RHG lattice.
The probability of logical error of the topological qubit depends on the probability that $XX$ and $ZZ$ measurement outcomes are erased.
A difference with the logical fusion analyses performed in previous sections is that now we need to differentiate an unsuccessful fusion due to gate failure, where only one of $XX$ and $ZZ$ is erased, with the unsuccessful case due to the loss, where both outcomes are erased. 
In fact, randomising the erased outcome in failed cases as in Ref.~\cite{fbqc}, which for general graph codes can be done via local Clifford operations, a failed fusion still has $50\%$ chance to provide the outcome required for either the primal or dual syndrome graph.   
Therefore, in unsuccessful cases, we seek to enhance logical failure instead of logical loss, leading to a different optimisation strategy.
To focus on a specific architecture, we consider the fault-tolerant fusion network constructed from fusing six-qubit hexagonal resource states from Ref.~\cite{fbqc} (see Fig.~\href{fig:FTconcatenations}{\ref{fig:FTconcatenations}a}), which has the highest measurement erasure threshold amongst the reported FBQC schemes, i.e. $12\%$. 
The loss threshold per photon is then the threshold at which $\overline{X}\overline{X}$ and $\overline{Z}\overline{Z}$ erasure probabilities are simultaneously suppressed below $12\%$. 
Such threshold in general also depends on the failure rate of the physical fusions employed, as trade-offs between boosting the fusion success rate and having to avoid loosing an increasing number of ancillary photons can be considered.
Such trade-offs result in concave curves for loss thresholds as a function of physical failure rates~\cite{fbqc}, as the ones shown in Fig.~\href{fig:FTconcatenations}{\ref{fig:FTconcatenations}c}.
In our analysis we optimise graphs considering only standard physical fusion gates, i.e. with $50\%$ success probability, but for completeness we report the performance also for boosted cases. 
By parallelising the optimisation procedure using the BlueCrystal high performance computing cluster, we optimise for graph states with up 
 to $n=10$ (i.e. 11 qubit progenitor graphs), considering both transversal and adaptive fusions.
The optimised graphs, shown in Appendix~\ref{sec::appendix_library}, provide the loss thresholds reported in Fig.~\href{fig:FTconcatenations}{\ref{fig:FTconcatenations}c}.
For the adaptive approach, the thresholds reach $10.5\%$ for qubit graph codes with $n=10$, considering non-boosted physical fusions, and $4.9\%$ for the transversal approach.
%


\section{Discussion}

We have shown how developing methods to analyse the measurement-based loss tolerance, as well as error correction properties, for arbitrary graph states can provide logical qubits with significantly higher noise tolerance and fewer physical qubits.
This is observed both for modules with $\simeq 10$ qubits and in the asymptotic regime where orders of magnitude improvements are observed with respect to tree graphs.
An immediate implication of these results is to show that most of the graph modules currently considered in various photonic-based applications, such as tree-based encodings for one-way and two-way quantum repeater protocols~\cite{Borregaard2020, Azuma2015, Hilaire2021resource, Zhan2022repeaterperf} and Shor-Beacon codes for logical fusions in FBQC architectures~\cite{fbqc}, are suboptimal.
Significant improvements can be obtained by using graphs optimised for the targeted functionality.
We illustrated these advantages for a few applications in Section~\ref{sec::applications}, but expect it to be relevant to improve a large part of photonic quantum applications based on graph states.
To this scope, we make the Python code utilised for all the analysis in this work freely accessible~\cite{githubTB}.
As an example, the per-photon-loss threshold of $10.5\%$ for fault-tolerant FBQC, obtained considering only standard non-boosted fusion gates, is a significant improvement with respect to the previous $2.7\%$ value with boosted fusions from Ref.~\cite{fbqc}, potentially bringing fault-tolerance much closer to the capabilities of near-term photonic hardware.
Moreover, this threshold is obtained considering the fusion network construction from Ref.~\cite{fbqc} based on fusing six-qubit hexagons as resource states, and we expect it to improve further by developing fusion networks with higher tolerance to fusion erasure~\cite{PaesaniBrown}.
Technologies that are in principle well suited for the generation of graph codes include
all-optical approaches, which when equipped with feed-forward and multiplexing could generate graph resource states deterministically~\cite{Browne2005, GimenoSegovia2015, bartolucci2021creation}, and approaches based on quantum emitters, where photonic entanglement can be directly generated via spin-photon interfaces~\cite{Lindner2009machinegun, Economou2010, gimeno2019}. 
In particular, high-fidelity spin-photon systems have been recently developed in a variety of platforms, including quantum dots, superconducting circuits, atoms in optical cavities, and NV centres~\cite{Appel2022, Tomm2021, Mart2022, Schwartz2016, Coste2022, Tiurev2022, Thomas2022}, with demonstrations of deterministic generation of graph states with up to 14 photons~\cite{Thomas2022}.
The tools developed here allowed us to identify loss-tolerant graph codes with minimal requirement in terms of number of qubits, and can be readily adapted to incorporate hardware-specific restrictions and error models.
We expect such capabilities to be significantly valuable in developing near-term experiments targeting loss tolerance in the photonic platform.
Such demonstrations will provide truly loss-tolerant photonic qubits, a milestone yet to be achieved that promise to unlock important opportunities for scaling photonic quantum technologies.
%
%


\section*{Acknowledgements}

We thank B. Brown, H. Shapourian, J. C. Adcock, A. E. Jones, B. Flynn, J. Borregaard, M. C. Löbl, A. S. Sørensen, and P. Lodahl for fruitful discussions.  
T.J.B. acknowledges support from UK EPSRC (EP/SO23607/1).
S.P. acknowledges funding from the Cisco University Research Program Fund (nr. 2021-234494) and from the Marie Skłodowska-Curie Fellowship project QSun (nr. 101063763).
Part of this work was carried out using the computational facilities of the Advanced Computing Research Centre, University of Bristol - http://www.bristol.ac.uk/acrc/.


\bibliographystyle{apsrev4-2}
\bibliography{biblio.bib}


\newpage 
\clearpage

\pagenumbering{arabic}

\onecolumngrid

\appendix

\setcounter{page}{1}

\renewcommand{\thesection}
{\Alph{section}}

\renewcommand{\thefigure}{\textbf{S\arabic{figure}}}
\renewcommand{\figurename}{\textbf{Supplementary Figure}}
\setcounter{figure}{0} 

\renewcommand{\thetable}{\textbf{S\arabic{table}}}
\renewcommand{\tablename}{\textbf{Supplementary Table}}
\setcounter{table}{0}

\counterwithout{equation}{section}
\renewcommand{\theequation}{S\arabic{equation}}
\setcounter{equation}{0}

\section{Decoders}
\label{sec:decoder}
\subsection{Loss-only decoding}

The decoding strategy for arbitrary basis measurements on graph codes implemented in this work is outlined in Algorithm \ref{alg:arbitrarydecoder}, with small modifications to that algorithm for Pauli basis or Fusion measurements.
The general strategy is as follows.
Firstly, all possible measurement are initialised.
For the case of Pauli measurements, this is simply the set of logical operators $\overline{\mathcal{L}}$, whereas for $\overline{A}(\theta)$ or fusion measurements, it is an operator satisfying the SPC~\cite{morley2019} on the progenitor graph for any choice of output qubit, combined with physical $A(\theta)$ or fusion measurements on that qubit.
Inspired by Ref.~\cite{morley2019} we construct measurements from only \textit{non-trivial} stabilizers and logical operators, i.e. those which cannot be decomposed in to a smaller weight operator multiplied by a stabilizer with non-overlapping support, reducing the number of operators to consider.
Then the optimal first measurement is determined according to a cost function - typically by choosing the measurement with the lowest weight, and selecting a random qubit from its support.
On attempting the measurement, the outcome is recorded ($\pm1$ or \textit{null} for a lost qubit), as well as the attempted basis.
Now the set of available measurements is updated in response to the outcome, according to Equation~\ref{eq:MBQEC_constraint}.
These three steps are repeated until a measurement has succeeded or no possible strategies remain.
Performing this search for every configuration of lost qubits constitutes building a decision tree for the decoder offline so that real-time decoding is simply a look-up table, which is queried up to $n$ times during measurement of an $n$-qubit graph code.
The tree is built using a depth-first search, and the termination conditions of the decoder mean that not all loss configurations need be examined.
Every leaf of the tree is an outcome of the decoding procedure, with an associated set of measured qubits $A$, lost qubits $B$, and an outcome $\in\{$\textit{success, fail}$\}$.
If we call the set of successful leaves as $Q$, the probability of a successful measurement for the graph state $G$ is
\begin{equation}
\label{eq::log_transmission_prob}
    \overline{\eta} = \mathcal{F}(\eta) = \sum_{q\in Q}P_{q}(\eta) = \sum_{q\in Q}\eta^{\abs{A_{q}}}(1-\eta)^{\abs{{B_{q}}}}
\end{equation}

This gives us analytic expressions for the effective transmission rates for various basis measurements on qubits encoded in graph codes.
\subsection{Loss and Unitary errors}
At a qubit level, mitigation of unitary errors is done by measurement of code qubits remaining after the target measurement has been completed - these are additional resources that we leverage to gain more information about the target measurement outcome. The decoder is accordingly adapted such that we no longer terminate after the loss-only decoder is finished, but instead determine which additional stabilizer measurements can be implemented to check the obtained outcome. When the optimal check measurements have been identified, they are attempted in succession, and as before after each the strategy is updated according to it's success or failure. This effectively extends the decision tree, reflecting the increased computational overhead of simultaneous error and loss correction. 

Choosing optimal check measurements is a non-trivial task. From the set of remaining valid stabilizers, a set of valid check operators must commute qubit-wise with one another, and with the set of measurements already performed. To choose a check set, we use heuristic methods, and pick the largest qubit-wise commuting set with the greatest overlap with the target measurement. It should be noticed that this is not necessarily optimal.

The error-tolerant performance of this approach is determined numerically. For each successful leaf of the decision tree, there is an associated target measurement, and a (possibly empty) set of check measurements. As detailed in the main text, we consider the phenomenological noise model, in which Pauli operators are randomly applied to each code qubit with probability $\lambda$. This may result in flipped measurement outcomes, which is a logical error on the measurement. To find the probability of a logical error, we consider all configurations of measurement error on all code qubits, which are binary strings of length $n$, denoted $\mathbf{e}$. Their probability is calculated, for the phenomenological model the probability of a flipped Pauli-basis measurement is $\epsilon_{\text{Pauli}} = 2\lambda$, as there are 2 anticommuting Pauli errors, and for an arbitrary basis measurement $\epsilon_{A} = 3\lambda$, so $P_{\mathbf{e}} = \sum_{i=1}^{n}\epsilon_{i}^{e_{i}}(1-\epsilon_{i})^{(1\oplus e_{i})}$. The syndrome is found from the outcomes of the check measurements, $\text{Synd}(\mathbf{e}) = \{
\pi(\mathbf{e}_{C_{j}})\}$, where $j$ indexes the $j$th check operator, and $\pi(x)$ denotes the parity of the string $x$. We also determine whether a logical error occurred on the target measurement, from the parity of $\mathbf{e}$ on the target measurement. From this, we can find the most probable error $P_{\text{max}}$ on the target given a particular syndrome, and so given a particular syndrome (which occurs with known probability) we correct for the most likely error pattern, succeeding with probability $P_{\text{max}}$. Summing over all syndromes gives the total error rate. This process needs to be done for each successful leaf of the decision tree, as the target and check measurements will differ, with the overall performance of the graph being the logical error probability for each leaf weighted by the probability of obtaining that configuration (Equation \ref{eq::log_transmission_prob}). Note that for logical arbitrary basis measurements we need to know the signs of two measurements to correctly decode the result, adding additional difficulty to the decoding process. Again, the computationally expensive parts of this decoding can be done offline, resulting in look-up table runtime costs.

\section{Adapting measurement patterns in Cascaded and Concatenated graphs}
\subsection{Cascaded graphs}
\label{sec::appendix_cascade}
Suppose we want to implement a local Pauli measurement pattern $M$ on a graph $G$, in order to implement either an indirect Pauli measurement or a SPF-based teleportation strategy. How do the required measurements change when another graph is appended to each qubit of $G$ (except the input)? The strategies outlined in this work are constructed from stabilizers, so we can consider how stabilizers from $G_{U}$ are modified by moving to the cascaded graph. By decomposing the stabilizer as outlined in Section~\ref{sec:graph states}, we see that if $b_{t}=0$, the modified measurement on the cascaded graph can have no weight on $G_{L}$, as $G_{L}$ is only adjacent to qubit $t$. $b_{t}=0$ additionally implies $M^{[t]} = Z$ or $\mathds{1}$, so $Z$ measurements do not need to be modified when switching to the cascaded graph. This is intuitive from the graphical perspective as well, the $Z_{t}$ measurement deletes that vertex from the graph, and $G_{L}$ is disconnected from $G_{U}$. For the same reasons, $\overline{Z}_{t}$ measurements can be performed by measuring qubits in $G_{L}$ only, leading to the improved loss-tolerance of cascaded graphs. If instead $b_{t}=1$ and $M^{[t]}=X$ or $Y$, the corresponding stabilizer in the cascaded graph penetrates in to $G_{L}$. The measurement pattern on the graph $G_{L}$ is $\prod_{j\in\mathcal{N}(t)}Z_{j}$, an indirect $X$ measurement on qubit $t$. Any X measurement on an intermediate layer code qubit therefore requires an additional indirect X basis measurement on deeper qubits, to disentangle them from the graph. This can be multiplied by any stabilizer of $G_{L} +t$ that does not include the generator $K_{t}(G_{L}+t)$, to ensure the measurements in $G_{U}$ are unaffected. If we multiply by an odd number of  stabilizer generators in the neighbourhood of $t$, this alters the required measurement on qubit $t$. For an arbitrary basis measurement, we can derive the measurement update rules by inspecting the changes to the stabilizers involved in constructing the measurement pattern (as outlined in Section~\ref{sec:arbitrary}). At least one of these stabilizers will include the generator of the output qubit, so in cascading this generator is modified as for $X$ or $Y$ measurements outlined above. These updates are then given by equation \ref{eq::cascade_update_rules}.

\begin{equation}
\label{eq::cascade_update_rules}
    \begin{split}
        X &\longrightarrow X_{t}\overline{X}_{t}(G_{L}+t) \text{ or } Y_{t}\overline{Y}_{t}(G_{L}+t) \\
        Y &\longrightarrow Y_{t}\overline{X}_{t}(G_{L}+t) \text{ or } X_{t}\overline{Y}_{t}(G_{L}+t) \\
        Z &\longrightarrow Z_{t} \mathds{1}_{G_{L}} \text{ or } \mathds{1}_{t} \overline{Z}_{t}(G_{L}+t) \\
        A &\longrightarrow A_{t}\overline{X}_{t}(G_{L}+t) \text{ or } \Tilde{A}_{t}\overline{Y}_{t}(G_{L}+t)
    \end{split}
\end{equation}

\begin{figure}[h!]
    \centering
    \includegraphics[scale=0.8]{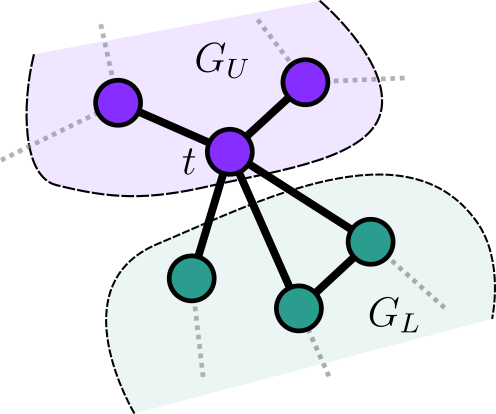}
    \caption{A Pauli measurement on an intermediate-layer qubit $t$ has additional requirements when we move to a cascaded graph. $t$ is a 'bottleneck' qubit between the upper and lower graphs $G_{U}$, $G_{L}$. We can find the properties of the casaded graph by considering the performance of the graphs $G_{U}$, $G_{L}+t$ for SPF and indirect Pauli measurements.}
    \label{fig::cascading_proof}
\end{figure}

Where $\Tilde{A}$ represents a modified basis. For the cases of $X$, $Y$ and $A$ the two options are incompatible, so one must choose which to attempt in advance. It is clear to see that both the direct and indirect measurements must succeed for one of these measurements to be successfully performed. The probability of successfully measuring a qubit in basis $M\neq Z$ becomes \begin{equation}
    \eta^{(k)}_{M, \text{casc}} = \eta \mathcal{F}(\mathbf{r}^{(k+1)};\;G_{k}, M)
\end{equation}
Where $\mathcal{F}(\mathbf{r}^{(k+1)};\;G_{k}, M)$ gives the probability of successfully performing a logical $M$ basis measurement as a function $\mathbf{r}^{(k+1)} = \{\eta^{(k+1)}_{X}, \eta^{(k+1)}_{Y}, \eta^{(k+1)}_{Z}, \eta^{(k+1)}_{A}\}$ on graph $G_{k}$ at depth $k$ in the cascade. $\eta$ is the physical transmission probability. This function can be calculated analytically by considering the small unit graphs that make up the cascade. For a $Z$ basis measurement, the two options in equation \ref{eq::cascade_update_rules} are compatible, so we can try both. The logical transmission is then calculated via
\begin{equation}
    \eta^{(k)}_{Z, \text{casc}} = \eta + (1-\eta) \mathcal{F}(\mathbf{r}^{(k+1)};\;G_{k}, Z)
\end{equation}

\subsection{Concatenated graphs}
\label{sec::appendix_concatenation}
In concatenated graphs the picture is similar, except now every every `bottleneck' qubit is a virtual qubit - it has been already measured in the $X$-basis, and the $+1$ outcomes obtained. The only measurement patterns than can be kept are then those compatible with $X_{t}$ measurements (using the same naming conventions as in Fig.~\ref{fig::cascading_proof}). For Pauli-basis measurements, these are simply read off from Equation \ref{eq::cascade_update_rules}. For the arbitrary basis measurement the situation is slightly different, as now we choose an output qubit in the deepest layer of the concatenation. As described in the main text, we can think of this as teleporting to a depth $k$ qubit, and then instead of measuring it in the arbitrary basis, teleporting it deeper in to the concatenation via the already performed X measurement of the virtual bottleneck qubits. Hence one can think of measuring the virtual qubit in an arbitrary basis by performing a teleportation measurement pattern on the qubits in the $k+1^{\text{th}}$ layer. 
\begin{equation}
    \begin{split}
        X &\longrightarrow X_{t}\overline{X}_{t}(G_{L} + t) \\
        Y &\longrightarrow X_{t}\overline{Y}_{t}(G_{L} + t) \\
        Z &\longrightarrow \overline{Z}_{t}(G_{L} + t) \\
        A &\longrightarrow \overline{A}_{t}(G_{L} + t)
    \end{split}
\end{equation}

It can be seen directly from these expressions that the effective transmission parameters for virtual qubits at depth $k$ is simply the probability of performing indirect measurements on qubits in the layer below, so can be recursively calculated according to the expression
\begin{equation}
    \eta_{M, \text{conc}}^{(k)} = \mathcal{F}(\mathbf{r}^{(k+1)};\;G_{k}, M)
\end{equation}

This form enables us to recover true loss-tolerance thresholds for MBQEC under code concatenation if the $\mathcal{F}$ are the same for all measurements in the target pattern. This is seen in the Steane code thresholds for MBQEC Pauli basis measurements discussed in the main text, where $\mathcal{F}(\mathbf{r};\;\text{Steane}, X) = \mathcal{F}(\mathbf{r};\;\text{Steane}, Y) = \mathcal{F}(\mathbf{r};\;\text{Steane}, Z)$.

\begin{figure}[]
    \centering
    \includegraphics[width=0.8\columnwidth]{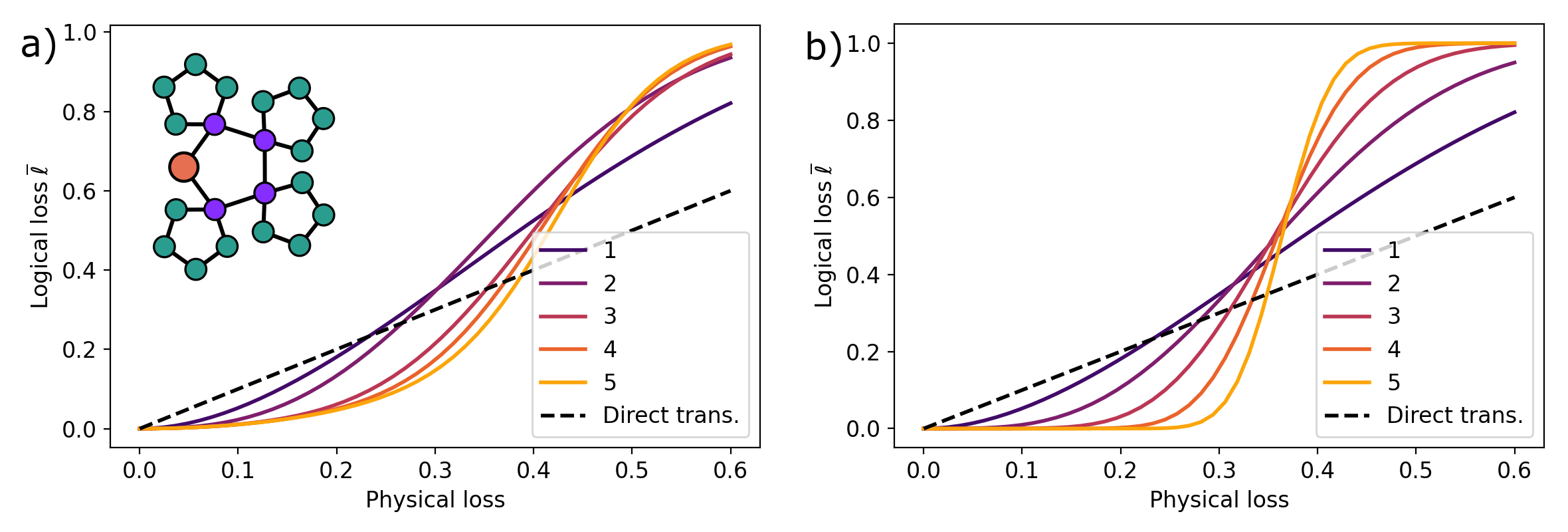}
    \caption{\textbf{Cascaded and Concatenated graphs}. a) Logical against physical loss rates for arbitrary measurements on cascaded pentagon graphs. Inset is a depth 2 cascade with 20 physical qubits (purple qubits are physical). b) shows the analogous performance for concatenated pentagons, for which the depth 2 concatenation has 16 physical qubits (purple qubits virtual).}
    \label{fig:cascaded_pentagon}
\end{figure}

In Fig.~\ref{fig:cascaded_pentagon} we compare cascaded and concatenated performance for an exemplary graph, here chosen to be the smallest progenitor graph with loss-tolerance for arbitrary basis measurements - the pentagon graph. 
Both methods suppress logical error rates, but we see that the concatenated graphs have lower loss rates and fewer code qubits.
In this work we have not considered how to physically realise these modular code constructions - it may be that the concatenated approach suffers from greater overhead during the preparation stage.
This would be an interesting and useful avenue for further investigations.

\section{Measurement complementarity principle}
\label{sec:complementarity}

By the uncertainty principle, one cannot perform measurements in different bases on a single qubit simultaneously, which has implications for the loss-tolerance thresholds of quantum codes, in particular that the thresholds may not exceed 50\%.
Consider an $[[n, 1, d]]$ stabilizer code, with stabilizer group $\mathcal{S}$ and logical operators $\overline{X}$, $\overline{Z}$.
If there exists a bipartition of the physical qubits into sets $A_{1}$, $A_{2}$ such that there exist logical operators whose support is restricted to each set $\Sigma(\overline{X}) \subseteq A_{1}$ and $\Sigma(\overline{Z}) \subseteq A_{2}$, the two logical operators would be simultaneously measurable, violating the uncertainty principle.
This shows that every pair of logical operators shares support on at least one qubit, $\Sigma(\overline{X}) \cap \Sigma(\overline{Z}) \neq \emptyset$.
One could imagine losing the set $A_{2}$ of physical qubits to a third party.
If either party is able to perform a measurement using only their qubits, the above argument necessitates that the other party can only recover the same measurement on their set.
To relate this to loss thresholds, we need to distinguish between the \textit{breakeven point} and the \textit{threshold}.
The break-even point refers to the point at which $\overline{\ell}(\ell_{b}) = \ell_{b}$ for a given code, whereas the threshold $l^{*}$ denotes the loss below which code concatenation increases the probability of measurement success.
Given a deep concatenation, the success probability in the subthreshold regime approaches 1.
The above argument places no restriction on the break-even point of the codes, but requires that for any two different measurements $P$ and $Q$, the loss-tolerance thresholds must satisfy $\ell^*[P] + \ell^*[Q] \leq 1$ - referred to as the measurement complementarity principle.
Hence, the threshold for arbitrary basis measurements cannot exceed 50\%.
This is an analogous restriction to the \textit{gate complementarity principle}~\cite{nickerson2018}, which applies to the probability of successfully performing logical gates.

\section{Local Clifford Equivalence}
\label{sec:LUE}
In this work we consider codes up to local Clifford operations, such that if two codes can be transformed in to each other by local Clifford operations they are deemed equivalent.
When searching for optimal loss tolerant codes we utilise the fact that the loss-tolerance of a graph code using the decoders implemented here is invariant under these operations.
This is seen by examining how graph codes are modified under local complementation, a graph transformation which inverts the neighbourhood of a particular node.
 Local complementation on a node $\alpha$ of a graph is equivalent to application of the following local unitary~\cite{VanDenNest2004,Hein2004PRA} to the the graph state.
\begin{equation}
    U_{\alpha}^{LC} = \sqrt{-iX_{\alpha}} \bigotimes_{\beta \in N_{\alpha}} \sqrt{iZ_{\beta}}
\end{equation}
where $N_{\alpha}$ denotes the neighbourhood of qubit $\alpha$.
Consider two graph states related via local complementation $\ket{G^{(2)}} = U_{\alpha}^{LC} \ket{G^{(1)}}$. 
These are the progenitor graphs of two locally-equivalent graph codes.
We will show here that the loss tolerance of these graph codes is identical.
Upon conjugation with $U_{\alpha}^{LC}$ the Pauli operators of a qubit $q$ in the graph transform as

\begin{center}
\begin{tabular}{ |c|c c c c| } 
 \hline
 & $X$ & $Y$ & $Z$ & $\mathds{1}$\\
 \hline
 $q = \alpha$ & $X$ & $-Z$ & $Y$ & $\mathds{1}$\\ 
 $q\in N_{\alpha}$ & $-Y$ & $X$ & $Z$ & $\mathds{1}$\\
  $q\not\in N_{\alpha}$ & $X$ & $Y$ & $Z$ & $\mathds{1}$\\
\hline
\end{tabular}
\end{center}

This transformation can be used to readily verify the stabilizer generators of $G^{(1)}$ transform in to the generators of $G^{(2)}$. 
The loss tolerance of $G^{(1)}$ under logical measurements in an arbitrary basis is determined by the set of valid measurement patterns $\mathcal{M} = \{M\}$, where $M = S^{(1)}_{1} \cup S^{(1)}_{2}$, such that the two stabilizers anticommute on the input and output, and commute on all other qubits.
Each stabilizer transforms under local complementation according to the above table, such that the qubit support is invariant, and any pair of operators which are (anti-)commuting on a particular qubit remain (anti-)commuting on that qubit.
This means $S^{(2)}_{1} \cup S^{(2)}_{2}$ is a valid measurement pattern on $G^{(2)}$, where $S^{(2)}_{j} = U_{\alpha}^{LC} S^{(1)}_{j} (U_{\alpha}^{LC})^{\dagger}$, and there is a bijective relation between measurements that perform measurement-based teleportation on graph states.
Furthermore, two measurement patterns that were initially compatible remain so, and the success probability is thus conserved. 

In general, a local Clifford operation preserves the bitwise commutativity of two Pauli operators, and the compatibility of two measurement patterns is determined by their bitwise commutation properties, so this argument applies similarly to measurements in Pauli bases, and to logical Fusion measurements.
For Pauli basis measurements, a local complementation of the progenitor graph may transform a logical Pauli operator in to a different logical Pauli, indeed the locally equivalent graphs depicted in Fig.~\ref{fig:Paulimeas} perform differently to one another in each basis.
However, the success probability will be conserved when averaging over Pauli bases, as all logical Paulis that are the same type in one graph will also be the same type in the locally-equivalent sibling.
The caveat to this invariance is in the decoder implementation. The decoder may make arbitrary choices between equally `good' measurement strategies, biasing the probability of performing measurements in particular bases.
A decoder based on heuristic methods may therefore not choose corresponding strategies in locally-equivalent graphs, so while their optimal loss tolerance is the same, their performance could vary in practice.

\section{Stabilized spaces under measurement and loss}
\label{sec:reduced_stab}
Consider a stabilizer code  with stabilizer $\mathcal{S}$ and logical operators $\overline{X}$ and $\overline{Z}$.
Suppose we perform a measurement $M \in \mathcal{P}^{n}$ on the qubits in the code.
The reduced stabilizer group is given by $\mathcal{S}_{M} = \{S \in \mathcal{S}\; |\; [S^{[i]}, M^{[i]}]=0\}$.
For any two stabilizers $S_{1}$, $S_{2}$ in the reduced group, their product $S_{3} = S_{1}S_{2}$ is also in the group. 
From the requirement that $[S^{[i]}, M^{[i]}]=0$, we obtain that $S^{[i]}= M^{[i]}$ or $\mathds{1}$.
The product $S_{1}^{[i]}S_{2}^{[i]} = M^{[i]}$ or $\mathds{1}$ from the properties of Pauli operators, and we obtain the commutation relations of $S_{3}$, $[S_{3}^{[i]}, M^{[i]}] = [S_{1}^{[i]}S_{2}^{[i]}, M^{[i]}] = 0$
The product element $S_{3}$ is therefore in the group, and it is closed. 
The identity element trivially remains in the group, and $-\mathds{1}$ cannot be added to the group by discarding elements, so the reduced set $\mathcal{S}_{M}$ forms a stabilizer group.
The effect of loss is similar, except for now we only retain stabilizers that act trivially on the lost set, i.e. if $M^{[i]}=\nexists$, $[S^{[i]}, M^{[i]}] \iff S^{[i]}=\mathds{1}$. 
From this it is straightforward to see, by using the same argument as above, that the restricted set of stabilizers satisfying this condition form a stabilizer group.

\section{Graph library}
\label{sec::appendix_library}

\begin{figure}[h!]
    \centering
\includegraphics[width=1\columnwidth]{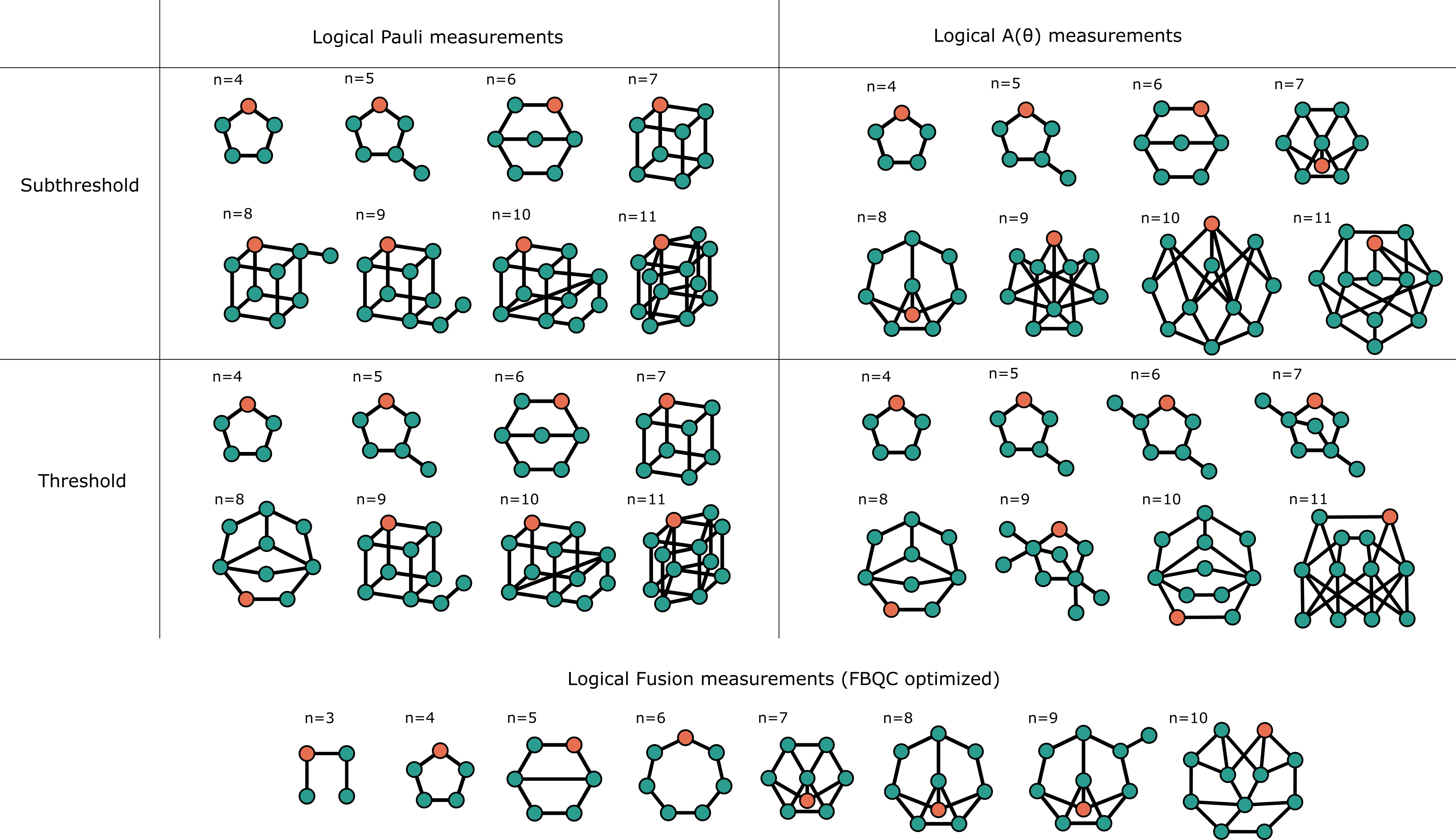}
    \caption{The optimised graphs for loss-tolerant logical qubit measurements (top), for near-threshold and subthreshold regimes, and optimised graphs for logical fusion measurements (bottom).}
    \label{fig:my_label}
\end{figure}

\end{document}